\documentclass[authoryear]{article}
\usepackage{setspace}
\usepackage{geometry}
\usepackage[utf8]{inputenc}
\usepackage{amsmath, amssymb, amsthm}
\usepackage[affil-it]{authblk} 
\usepackage{natbib}
\usepackage{multirow}
\usepackage{ulem}
\usepackage{xcolor}
\usepackage{graphicx}
\usepackage{subfig}
\usepackage{algorithm}
\usepackage{algpseudocode}
\RequirePackage[colorlinks,citecolor=blue,urlcolor=blue]{hyperref}
\graphicspath{ {./figures} }

\def\spacingset#1{\renewcommand{\baselinestretch}%
{#1}\small\normalsize} \spacingset{1}
\usepackage{float}

\title{Flexible Method Comparison with the Probability of Agreement}

\author{Nathaniel T. Stevens \\ Department of Statistics and Actuarial Science, University of Waterloo}

\date{}

\begin{document}
\newcommand{\todo}[2][red]{\footnote{\textcolor{#1}{#2}}}

\maketitle

\begin{abstract}
 The comparison of methods of measurement is a common problem in clinical practice; as novel methods are developed, establishing their agreement with existing methods is crucial. The probability of agreement (PoA) has previously been proposed as an intuitive and informative means of assessing agreement between two methods of measurement. It straightforwardly quantifies the likelihood that two measurements by different methods on the same subject are clinically indistinguishable. In this paper, we overhaul and extend the PoA methodology by developing an inference framework that relaxes several restrictive assumptions made in previous implementations, ultimately increasing its utility in a wider range of applications. We illustrate this more flexible methodology in an example that compares methods of measuring total Prostatic Specific Antigen (tPSA). And we thoroughly investigate its performance via simulation. This work dramatically increases the flexibility, availability, and hence impact of the PoA approach for method comparison.

\vspace{5mm}

\end{abstract}

\spacingset{2} 
\newpage

\section{Introduction} \label{sec:intro}

The comparison of methods of measurement is a common problem in clinical practice; as novel methods are developed, establishing their agreement---and hopefully interchangeability---with existing methods is crucial. So called method comparison studies are routinely undertaken to assess the comparability of new and existing methods, and the design and analysis of such studies has garnered much attention in medical research. For instance, one of the seminal works in this area, the \citet{bland1986statistical} \textit{Lancet} paper, is among the most highly cited statistics papers of all time \citep{ryan2005most} and was named the 29th most highly cited paper ever, across all disciplines \citep{mansournia2021bland}. While the \textit{limits of agreement} approach proposed in this paper and elaborated upon in others \citep{altman1983measurement, bland1986statistical,bland1995comparing, bland1999measuring, bland2003applying, bland2007agreement} remains widely popular, many other contemporary statistical techniques have been developed to assess agreement; see \citet{parker2020using} and \citet{brousseau2026scoping} for recent reviews and illustrations. 

The present paper considers one such contemporary development: the \textit{probability of agreement} (PoA). The PoA was proposed as an intuitive and informative alternative to the limits of agreement methodology \citep{stevens2017assessing,stevens2018comparing}; it straightforwardly quantifies the likelihood that two measurements by different methods on the same subject are clinically indistinguishable. Defining $\delta$ to be an equivalence margin, the interval $(-\delta,\delta)$ represents a region of clinical equivalence; differences within this interval are practically negligible, and therefore, what \citet{bland1986statistical} term, \textit{clinically acceptable}. The PoA is formally defined in this setting as 
\begin{equation}\label{eq:poa1}
\text{PoA}(s)=\text{Pr}(-\delta\leq Y-X \leq \delta|s),
\end{equation}
where $X$ and $Y$ respectively denote measurements taken by a reference and comparison method on a subject whose true underlying latent trait being measured has value $s$; it is the probability that the difference in measurements by the two methods on such subjects is clinically acceptable. Thus, values of this probability close to 1 suggest the methods are in agreement, and values close to 0 suggest disagreement. This probability can be calculated across a range of $s$ values of interest, and therefore can be used to make judgments about method agreement that depend on the quantity being measured; it may be the case that agreement is heterogeneous and two methods agree for some subset of the population and not others. 

Since its initial proposal \citep{stevens2017assessing}, the use of the PoA for method comparison has been extended in several directions. For instance, \citet{stevens2018comparing} relaxed an assumption of homoscedastic measurement errors, \citet{de2021bayesian} developed a Bayesian extension meant to accommodate the simultaneous comparison of several measurement methods, and \citet{ahmadi2024assessing} extended the methodology to jointly account for repeatability and reproducibility in settings for which the measurement method is used by multiple operators. Beyond method comparison, the PoA has proved to be a broadly useful methodology for quantifying agreement between statistical quantities in a variety of applications. For instance, PoA methods have been developed for the comparison of surgical learning curves \citep{ahmadi2026risk}, different arms in online A/B tests \citep{stevens2022comparative}, survival distributions \citep{stevens2020bayesian, stevens2020comparing}, spatial variables \citep{acosta2024comparing}, team performances in data competitions \citep{anderson2019host}, as well as for purposes of model validation \citep{ledwith2023probabilities}.

Recently, \citet{taffe2023use} criticized the original implementations of the probability of agreement \citep{stevens2017assessing,stevens2018comparing} for making the restrictive assumption that the true latent trait follows a normal distribution. To overcome this limitation, he proposed an estimation framework for the PoA that purportedly does not require a distributional assumption for the latent trait. However, despite what is claimed in the paper, implementations of the estimation framework in both his \texttt{MethodCompare} package in R \citep{taffe2019methodcompare} and his \texttt{ctl} package in Stata \citep{taffe2025ctl} explicitly rely on an assumption of normality for the underlying trait\footnote{In both implementations, the best linear unbiased predictor (BLUP) for the latent trait is estimated by fitting a linear mixed-effects model. Both implementations---\texttt{lme4::lmer} in R \citep{bates2015fitting} and \texttt{mixed} in Stata \citep{statacorp2013stata}---fit the model using maximum likelihood (or restricted maximum likelihood) assuming the latent trait's random effect is normally distributed.}. Moreover, the proposed procedure for interval estimation, a parametric bootstrap, also explicitly assumes the latent trait is normally distributed. In the present paper we \textit{truly} overcome the need to assume normality.

In particular, this paper contributes to clinical practice by making available a version of the PoA methodology that relaxes several restrictive assumptions made in previous implementations, ultimately increasing its utility in a wider range of applications. Specifically, the extension developed here (i) removes the need for the aforementioned normality assumption; (ii) allows for potential non-linearity in bias and precision; (iii) accommodates a non-constant equivalence margin; and (iv) does not require a balanced study design. Additionally, we provide freely-available and user-friendly R code to implement the proposed analyses. This work dramatically increases the flexibility, availability, and hence impact of the PoA approach for method comparison.

The remainder of the paper is organized as follows. In Section \ref{sec:methods} we describe the extended, more flexible, version of the PoA developed in this paper and in Section \ref{sec:ex} we demonstrate its use in the context of a study designed to compare methods of measuring total Prostatic Specific Antigen (tPSA). In Section \ref{sec:sim} we explore, via extensive simulation, the performance of the proposed methodology to broadly build confidence in its application. And we conclude with a summary and discussion of future work in Section \ref{sec:conc}. Various mathematical details and additional simulation results are relegated to Appendices \hyperref[sec:appxA]{A}, \hyperref[sec:appxB]{B}, \hyperref[sec:appxC]{C} and \hyperref[sec:appxD]{D}, and the code that accompanies the paper is available on GitHub at: \url{https://anonymous.4open.science/r/PoA4FlexMC-1E68}.

\section{The Probability of Agreement} \label{sec:methods}
We consider the comparison of two methods of measurement in the context of a comparison study in which both methods measure multiple subjects multiple times each. The goal, with the data collected, is to use the probability of agreement to quantify the level of agreement between the measurements made by the two methods. In the subsections that follow, we define the probability of agreement, the requisite model, and we describe estimation, inference, and other issues relevant to the practical application of the methodology.

\subsection{The Model} \label{sec:model}
 In order to define and calculate the probability of agreement, we first require a model that describes the data observed in the comparison study. The model we use is similar to those considered by \citet{stevens2017assessing,stevens2018comparing, taffe2019methodcompare, taffe2020assessing, taffe2023use, taffe2025ctl} but, importantly, we relax potentially restrictive distributional assumptions as well as assumptions concerning the relative bias between the methods and the methods' precisions. In particular, we accommodate a non-linear relative bias as well as non-constant and potentially non-linear precisions. The model is as follows: 
\begin{equation}\label{eq:model}
\begin{aligned}
    X_{ij}|(S_i=s) &= s + M_{x,ij}, &\text{where}~~M_{x,ij}|(S_i=s) \sim \mathcal{N}\left(0, \sigma^2_x(s;\boldsymbol{\omega}_x)\right)\\
    Y_{ij}|(S_i=s) &= g(s;\boldsymbol{\beta}) + M_{y,ij}, &\text{where}~~ M_{y,ij}|(S_i=s) \sim \mathcal{N}\left(0, \sigma^2_y(s;\boldsymbol{\omega}_y)\right)
\end{aligned}
\end{equation}
\noindent where $i=1,2,\ldots,n$ indexes subjects and $j=1,2,\ldots,r_{x,i}$ (or $r_{y,i}$) indexes the replicate measurements made on subject $i$ by method $X$ (or $Y$). We emphasize that the number of replicates need not be the same for each method-subject combination, but we do require that $r_{x,i}\geq2$ and $r_{y,i}\geq2$ for all $i$. That is, we require \textit{some} replication on each subject by each method. 

We let $S_i$ denote the true underlying latent trait being measured. We do not need to specify a distribution for $S_i$, we need only assume it has finite mean and variance, $\text{E}[S_i]=\mu$ and $\text{Var}[S_i]=\sigma^2$. Defined in this way, the model assumes the references method $X$ measures without bias, and it accommodates a possible non-linear bias $g(s;\boldsymbol{\beta})$ in the comparator method $Y$'s measurements. While many techniques may be used to flexibly model such a non-linear relationship, we use a polynomial of the form $$g(s;\boldsymbol{\beta})=\beta_0+\beta_{1}s +\cdots+\beta_{p}s^p.$$ 

We denote the measurement error inherent to each method by $M_{x,ij}$ and $M_{y,ij}$. These random variables are assumed to be normal, mean zero, and potentially heteroscedastic; the model allows each method's measurement variability (precision) to depend non-linearly on the size of the true underlying characteristic being measured via the functions $\sigma_x(s;\boldsymbol{\omega}_x)$ and $\sigma_y(s;\boldsymbol{\omega}_y)$. As with the non-linear bias, we model non-linear precision via polynomials of the form
\begin{align*}
\sigma_x(s;\boldsymbol{\omega}_x) &= \omega_{x,0}+\omega_{x,1}s +\cdots+\omega_{x,d_x}s^{d_x} \\
\sigma_y(s;\boldsymbol{\omega}_y) &= \omega_{y,0}+\omega_{y,1}s +\cdots+\omega_{y,d_y}s^{d_y}
\end{align*}  
We further assume that $M_{x,ij}$ and $M_{y,ij}$ are independent of each other and themselves for all $(i,j)$. The independence and normality assumptions made for these errors are routine, and can be assessed graphically with QQ-plots, histograms, and scatterplots of residuals. 

The proposed model and estimation procedure requires a choice for the the polynomial orders $(p, d_x, d_y)$, and we recommend that this choice be data-driven. In particular, we recommend fitting polynomials of several orders and selecting the order that yields the best fit to the data, which we operationalize as the one which minimizes unexplained variation as quantified by the Bayesian information criterion $$\widehat{BIC} = n\log\left(\frac{RSS}{n}\right)+k\log(n),$$ where $RSS$ is the residual sum of squares associated with the fitted polynomial and $k$ is the number of coefficients in the fitted polynomial. Acknowledging that this value is subject to sampling variation, one may instead choose the optimal polynomial order based on the bootstrap bias-corrected estimate of $BIC$ \citep{hesterberg2015teachers} given by $2\widehat{BIC}-\overline{BIC^*}$, where $\overline{BIC^*}$ is the average of $B$ estimates of $BIC$ obtained from $B$ bootstrap resamples of the data used to fit the relevant polynomial. Of course, residual diagnostics and significance testing may also support order selection, but we investigate the performance of the $BIC$-based selection procedure in Section \ref{sec:sim}.

\subsection{The  Metric}\label{sec:poa}
With this model, we can define the probability of agreement from Equation \eqref{eq:poa1} more precisely:
\begin{equation} \label{eq:poa2}
\begin{aligned}
\text{PoA}(s) &= \text{Pr}\left(-\delta \leq Y_i-X_i \leq \delta | S_i = s\right) \\ 
& = \Phi\left(\frac{\delta-g(s;\boldsymbol{\beta})+s}{\sqrt{\sigma_x^2(s;\boldsymbol{\omega}_x) + \sigma_y^2(s;\boldsymbol{\omega}_y)}}\right) - \Phi\left(\frac{-\delta-g(s;\boldsymbol{\beta})+s}{\sqrt{\sigma_x^2(s;\boldsymbol{\omega}_x) + \sigma_y^2(s;\boldsymbol{\omega}_y)}}\right) 
\end{aligned}
\end{equation}
where $\Phi(\cdot)$ is the standard normal cumulative distribution function arising because $Y_i-X_i|(S_i=s) \sim \mathcal{N}\left(g(s;\boldsymbol{\beta})-s ~, ~ \sigma_x^2(s;\boldsymbol{\omega}_x) + \sigma_y^2(s;\boldsymbol{\omega}_y)\right)$, which is itself a consequence of the assumed normality of the measurement errors. 

The PoA in Equation \eqref{eq:poa2} is estimated by plugging in estimates of the parameters $\boldsymbol{\beta}$, $\boldsymbol{\omega}_x$, and $\boldsymbol{\omega}_y$ (whose estimation is discussed in Section \ref{sec:est}):
\begin{equation}\label{eq:poa-est}
\widehat{\text{PoA}}(s) = \Phi\left(\frac{\delta-g(s;\hat{\boldsymbol{\beta}})+s}{\sqrt{\sigma_x^2(s;\hat{\boldsymbol{\omega}}_x) + \sigma_y^2(s;\hat{\boldsymbol{\omega}}_y)}}\right) - \Phi\left(\frac{-\delta-g(s;\hat{\boldsymbol{\beta}})+s}{\sqrt{\sigma_x^2(s;\hat{\boldsymbol{\omega}}_x) + \sigma_y^2(s;\hat{\boldsymbol{\omega}}_y)}}\right).
\end{equation}
Values of this quantity near 1 (0) suggest that it is (un)likely that measurements by methods $X$ and $Y$ on a subject with latent trait $s$ are clinically equivalent (i.e., they differ by an amount that is clinically acceptable). Acknowledging that $\widehat{\text{PoA}}(s)$ depends on both $\delta$ and $s$, we make the following remarks. 

\noindent \textbf{Remark 1:} The equivalence margin $\delta$ critically impacts the assessment of agreement; larger (smaller) values of $\delta$ necessarily yield larger (smaller) $\widehat{\text{PoA}}(s)$ values. As such, $\delta$ should be defined carefully by subject matter experts to facilitate a meaningful comparison. Additionally, $\delta$ need not be a constant, nor must the equivalence interval be symmetric around 0. It is trivial to extend the PoA with $(-\delta,\delta)$ replaced by $(\delta_L(s), \delta_U(s))$, where $\delta_L$ and $\delta_U$ are possibly asymmetric functions of the true underlying characteristic $s$. We illustrate this in Section \ref{sec:ex} and Appendix \hyperref[sec:appxB]{B}.

\noindent \textbf{Remark 2:} Because $\widehat{\text{PoA}}(s)$ depends on $s$, we propose constructing a \textit{probability of agreement plot}, a plot of $\widehat{\text{PoA}}(s)$ vs. $s$, which dynamically displays agreement as a function of the underlying characteristic being measured. It is often the case that two methods may agree for some values of the latent trait and not others, and such a plot aids in this assessment. In recommending this plot, it is important to acknowledge that we do not (and cannot) know the true value of the latent trait $s$; we simply recommend calculating (and plotting) Equation \eqref{eq:poa-est} across a range of $s$ values of practical interest. Alternatively, we could estimate $s$ for a given subject (yielding $\hat{s}$) and calculate $\widehat{\text{PoA}}(\hat{s})$. We describe this approach and its implications in Section \ref{sec:condpoa}, and we illustrate these various PoA calculations in Section \ref{sec:ex}.

\subsection{Point Estimation} \label{sec:est}

In order to estimate the PoA as in Equation \eqref{eq:poa-est}, we must first estimate the model parameters $\boldsymbol{\beta}$, $\boldsymbol{\omega}_x$, and $\boldsymbol{\omega}_y$ given the observed measurements $(x_{i1},x_{i2}, \ldots, x_{ir_{x,i}})$ and $(y_{i1},y_{i2}, \ldots, y_{ir_{y,i}})$ for each subject ($i=1,\ldots,n$). Recognizing that model \eqref{eq:model} is a non-linear errors-in-variable regression model, we adopt the regression calibration\footnote{Note: \textit{regression} calibration is not related to \textit{measurement method} calibration discussed at the end of this section.} approach to parameter estimation and begin by estimating $s_1,\ldots,s_n$, the underlying latent trait for each of the $n$ subjects. We then estimate $\boldsymbol{\beta}$, $\boldsymbol{\omega}_x$, and $\boldsymbol{\omega}_y$ using the estimates $\hat{s}_1,\ldots,\hat{s}_n$. 

The quantity $s_i$ is estimated by the best linear approximation $\hat{s}_i$ found by modeling the unobserved $S_i$ as a function of the measurements made by the reference method $X$ \citep{carroll2006measurement}. In particular, we have 
\begin{equation}\label{eq:bla}
\hat{s}_i = \hat{\mu}+\frac{\hat\sigma^2(\overline{x}_i-\hat\mu)}{\hat\sigma^2+\hat\sigma_{x,i}^2/r_{x,i}}    
\end{equation}
where $\hat\mu$, $\hat\sigma^2$, and $\hat\sigma_{x,i}^2$ are estimates of $\mu$, $\sigma^2$, and $\sigma_{x,i}^2$ based on the references method's measurements $(x_{i1},x_{i2}, \ldots, x_{ir_{x,i}})$, $i=1,\ldots,n$. Note $\sigma_{x,i}^2$ quantifies the precision of the reference method for subject $i$ specifically. These estimates are calculated as follows:
\begin{align*}
\hat\mu &= \frac{1}{N}\sum_{i=1}^n\sum_{j=1}^{r_{x,i}}x_{ij} \\
\hat\sigma^2_{x,i} &= \frac{1}{r_{x,i}-1}\sum_{j=1}^{r_{x,i}}(x_{ij}-\overline{x}_i)^2 \\
\hat\sigma^2 &= \frac{\sum_{i=1}^{n}r_{x,i}\left(\overline{x}_i-\overline{x}\right)^2-\sum_{i=1}^n(1-r_{x,i}/N)\hat\sigma^2_{x,i}}{N-\frac{1}{N}\sum_{i=1}^nr^2_{x,i}}
\end{align*}
where $\overline{x}_i=r_{x,i}^{-1}\sum_{j=1}^{r_{x,i}}{x_{ij}}$ is the mean of replicate measurements on subject $i$ by the reference method, and $\overline{x}=N^{-1}\sum_{i=1}^n\sum_{j=1}^{r_{x,i}}{x_{ij}}$ is the overall average of all $N=\sum_{i=1}^n r_{x,i}$ measurements taken by the reference method. The legitimacy of these formulae are explained in Appendix \hyperref[sec:appxA]{A}.

With $(\hat{s}_1,\ldots,\hat{s}_n)$ determined, these values are inputted into the estimation procedures for $\boldsymbol{\beta}$, $\boldsymbol{\omega}_x$, and $\boldsymbol{\omega}_y$, the coefficients that define the bias and precision functions $g(s;\boldsymbol{\beta})$, $\sigma_x(s;\boldsymbol{\omega}_x)$, and $\sigma_y(s;\boldsymbol{\omega}_y)$. In particular, we estimate $\boldsymbol{\omega}_x$ and $\boldsymbol{\omega}_y$ via polynomial regressions relating the within-subject standard deviations $\hat\sigma_{x,i}$ and $\hat\sigma_{y,i}$ to the best linear approximation $\hat{s}_i$. For each method separately, we solve the following least squares objective: $$\hat{\boldsymbol{\omega}} = \operatorname*{argmin}_{\boldsymbol{\omega}\in\mathbb{R}^{d+1}} ~\sum_{i=1}^n\left(\hat\sigma_i - \omega_{0}-\omega_{1}\hat{s}_i -\cdots-\omega_{d}\hat{s}_i^{d}\right)^2.$$ Next, we estimate the bias coefficients by solving the following weighted least squares objective: $$\hat{\boldsymbol{\beta}} = \operatorname*{argmin}_{\boldsymbol{\beta}\in\mathbb{R}^{p+1}} ~\sum_{i=1}^n\sum_{j=1}^{r_{y,i}}\frac{\left(y_{ij} - \beta_{0}-\beta_{1}\hat{s}_i -\cdots-\beta_{p}\hat{s}_i^{p}\right)^2}{\hat\sigma_y^2(\hat{s}_i;\hat{\boldsymbol{\omega}}_y)}.$$ With $\hat{\boldsymbol{\beta}}$, $\hat{\boldsymbol{\omega}}_x$, and $\hat{\boldsymbol{\omega}}_y$ determined in this way, the PoA estimate in Equation \eqref{eq:poa-est} can be calculated. In Section \ref{sec:ex}, we illustrate the use of this estimation procedure, and in Section \ref{sec:sim} we investigate its performance more broadly.

In addition to PoA estimation, another use of $\boldsymbol{\beta}$, $\boldsymbol{\omega}_x$, and $\boldsymbol{\omega}_y$ estimates is \textit{method calibration}. If there exists a significant relative bias between the comparator and reference methods, the comparator measurements can be de-biased and hence calibrated. By model \eqref{eq:model}, we have that $\text{E}[Y_{ij}|S_i=s]=g(s;\boldsymbol{\beta})$. So, given an observed measurement $y_{ij}$ and estimated bias parameters $\hat{\boldsymbol{\beta}}$, we define the corresponding calibrated measurement as $y_{ij}^c=g^{-1}(y_{ij};\hat{\boldsymbol{\beta}})$ where we $g^{-1}$ is determined numerically. Note that for some polynomial orders $d$ and estimates $\hat{\boldsymbol{\beta}}$, $g^{-1}$ may not exist in the domain of $s$ considered. In such cases, calibration is unavailable and perhaps a linear (or at least lower-order) polynomial should be considered instead for $g(s;\boldsymbol{\beta})$. We illustrate the benefit of a calibrated agreement analysis in Section \ref{sec:ex}.

\subsection{Interval Estimation} \label{sec:inf}

Denote the parameter vector of interest as $\boldsymbol{\theta}=(\mu,\sigma,\boldsymbol{\beta}^\top, \boldsymbol{\omega}_x^\top, \boldsymbol{\omega}_y^\top)^\top\in\mathbb{R}^m$. For flexible inference that does not rely on any distributional assumptions, we use the bootstrap to construct confidence intervals for the various model parameters. To preserve the dependence among replicate measurements on a given subject, we specifically employ \textit{cluster bootstrapping} \citep{sherman1997comparison,cheng2013cluster} in which the $n$ subjects are resampled with replacement and all replicate measurements (by each method) are included in the resulting bootstrap sample. Constructing $B$ bootstrap samples in this way yields $B$ bootstrap estimates of the parameters $\hat{\boldsymbol{\theta}}^*_1,\hat{\boldsymbol{\theta}}^*_2,\ldots,\hat{\boldsymbol{\theta}}^*_B$, from which bootstrap confidence intervals are constructed. We consider both standard (Section \ref{sec:ci_stand}) and percentile-based (Section \ref{sec:ci_quant}) bootstrap intervals \citep{efron2021computer}. For the PoA, we also consider both pointwise confidence intervals as well as simultaneous confidence bands. In Section \ref{sec:sim} we investigate the coverage of these various bootstrap interval estimates.

\subsubsection{Standard Bootstrap Confidence Intervals} \label{sec:ci_stand}
The \textit{standard} $(1-\alpha)\times100\%$ bootstrap confidence interval for an individual parameter $\theta$ is given by $\hat\theta\pm z_{1-\alpha/2}\text{SE}_{boot}[\hat\theta]$ where $\text{SE}_{boot}[\hat\theta]$ is the empirical standard deviation of the $B$ bootstrap estimates $\hat\theta^*_1, \hat\theta^*_2, \ldots, \hat\theta^*_B$. This is the method employed to construct CIs  for the individual parameters in $\boldsymbol{\theta}$ associated with model \eqref{eq:model}. Confidence intervals for $\text{PoA}(s)$ could naively be constructed in the same way, but this would not guarantee the intervals are contained within $[0,1]$. To ensure CIs for $\text{PoA}(s)$ obey this constraint we first transform the bootstrap PoA estimates using a differentiable and monotonically increasing function $f:[0,1]\rightarrow\mathbb{R}$, construct the standard CI on the transformed scale, and then back-transform those confidence limits (using $f^{-1}$), yielding an interval for $\text{PoA}(s)$. While many functions $f$ may be used, we use the complementary log-log transformation $f(p)=\log(-\log(1-p))$.

Because $\text{PoA}(s)$ is likely to be estimated (and plotted) along a continuum of $s$ values, it may be preferable to construct a simultaneous confidence band for the whole curve, as opposed to pointwise intervals for specific $s$ values. We do so by employing the delta method framework described in \citet{cheng2005bootstrapping}. In particular, a $(1-\alpha)\times100\%$ simultaneous confidence band for some function $\eta(s;\boldsymbol{\theta})$ is given by $[L(s),U(s)]=\eta(s;\hat{\boldsymbol{\theta}})\pm h(s;\hat{\boldsymbol{\theta}})$ where $$h(s;\hat{\boldsymbol{\theta}})=\sqrt{\chi^2_{m}(1-\alpha)\left[\frac{\partial\eta(s;\boldsymbol{\theta})}{\partial\boldsymbol{\theta}}\right]^\top_{\boldsymbol{\theta}=\hat{\boldsymbol{\theta}}}V\left(\hat{\boldsymbol{\theta}}\right)\left[\frac{\partial\eta(s;\boldsymbol{\theta})}{\partial\boldsymbol{\theta}}\right]_{\boldsymbol{\theta}=\hat{\boldsymbol{\theta}}}},$$ and $\chi^2_{m}(1-\alpha)$ is the $1-\alpha$ quantile of the $\chi^2_{m}$ distribution with $m$ degrees of freedom, and $V\left(\hat{\boldsymbol{\theta}}\right)$ is the sample variance-covariance matrix calculated from the $B$ bootstrap estimates $\hat{\boldsymbol{\theta}}^*_1,\hat{\boldsymbol{\theta}}^*_2,\ldots,\hat{\boldsymbol{\theta}}^*_B$. Because we're constructing a confidence band for the PoA, we define $\eta(s;\boldsymbol{\theta})=f(\text{PoA}(s))$ with the complementary log-log transformation $f$ as above. The resulting confidence band for the PoA is then defined with $\left[f^{-1}(L(s)), f^{-1}(U(s))\right]$. The relevant partial derivatives needed for this calculation are provided in Appendix \hyperref[sec:appxB]{B}. Note that analogous confidence bands can also be constructed for the bias and precision functions $g(s;\boldsymbol{\beta}),\sigma_x(s;\boldsymbol{\omega}_x),\sigma_y(s;\boldsymbol{\omega}_y)$. Appendix \hyperref[sec:appxB]{B} provides the details of those calculations as well.

\subsubsection{Percentile-Based Bootstrap Confidence Intervals} \label{sec:ci_quant}

The confidence interval (and band) methods described in the previous subsection rely on the assumption that the sampling distribution of an estimate is normally distributed. While often valid, in certain circumstances this assumption may not be appropriate and the resulting performance (i.e., coverage) of the associated confidence intervals may suffer. As a fully nonparametric alternative, here we consider bootstrap-based confidence intervals whose construction is rooted in the \textit{percentile method}. A $(1-\alpha)\times100\%$ confidence interval for an individual parameter $\theta$ is trivially defined as the middle $(1-\alpha)\times100\%$ of the corresponding bootstrap distribution. That is, the $\alpha/2$ and $1-\alpha/2$ percentiles of the $B$ bootstrap estimates $\hat\theta^*_1, \hat\theta^*_2, \ldots, \hat\theta^*_B$: $\left(\hat\theta_{[\%\text{ile}]}(.025),\hat\theta_{[\%\text{ile}]}(.975)\right)$. 

This method works satisfactorily for the individual parameters in model \eqref{eq:model} as well as pointwise CIs for $\text{PoA}(s)$, but extra care needs to be taken if interest lies in constructing a percentile-based simultaneous confidence band for the PoA curve; the notion of percentiles must be extended to the collection of $B$ bootstrap curves $\widehat{\text{PoA}}_1^*, \widehat{\text{PoA}}_2^*, \ldots, \widehat{\text{PoA}}_B^*$. We do so using the band depth approach where a ``percentile'' curve is determined using \textit{modified band depths} \citep{lopez2009concept,lopez2010robust}. The calculation is as follows $$MBD\left(\widehat{\text{PoA}}_b^*\right) = {B \choose2}^{-1}\sum_{j<k}\frac{1}{|\mathcal{S}|}\sum_{s\in\mathcal{S}}\mathbb{I}\left\{\min(\widehat{\text{PoA}}_j^*(s),\widehat{\text{PoA}}_k^*(s)) \leq \widehat{\text{PoA}}_b^*(s)\leq\max(\widehat{\text{PoA}}_j^*(s),\widehat{\text{PoA}}_k^*(s))\right\},$$ where $\mathcal{S}$ is a set of $s$ values the PoA is calculated over, and $|\mathcal{S}|$ is the cardinality of that set. This measures the average proportion of the curve $\widehat{\text{PoA}}_b^*$ that is contained within any two other bootstrap PoA curves. The calculation is performed for \textit{all} $B$ bootstrap PoA curves and the confidence band is taken to be the convex hull of the middle $(1-\alpha)\times100\%$ of the bootstrap curves. That is, those with the $(1-\alpha)\times100\%$ largest MBD values.  Although no distributional assumption is required here, the tradeoff is computational complexity; in addition to the resampling needed for the bootstrap, the MBD calculation is computationally intensive: $\mathcal{O}(B^2\times|\mathcal{S}|)$.

\subsection{Conditional Probability of Agreement} \label{sec:condpoa}

The PoA as described above is calculated assuming inference about method agreement in a \textit{population} of individuals is of interest. However, in certain circumstances, interest may lie in assessing the agreement between methods for \textit{specific individuals}. In this case, the ``conditional'' probability of agreement suggested by \citet{taffe2023use} may be of interest. However, the Taffé methodology relies on a normal assumption for the true underlying characteristic, and it also yields the same PoA estimate for two individuals if they have the same measurements by method $X$, no matter what their measurements by method $Y$ might be. We propose here a different implementation of the same concept that overcomes these limitations.

The conditional PoA estimate for a given subject is calculated as in equation \eqref{eq:poa-est}, except that we plug in that subject's estimate of $s$ instead of a generic value of $s$ of interest. Thus, the conditional PoA for subject $i$ is $\widehat{\text{PoA}}(s=\hat{s}_i)$, $i=1,2,\ldots,n$.  Note that \citet{taffe2023use} calculates $\hat{s}$ using measurements from the reference method only. Doing so yields the same conditional PoA estimate for two subjects with the same $X$ values, even if they have drastically different $Y$ values. To obtain a truly subject-specific conditional PoA estimate, we propose calculating $\hat{s}$ as in Equation \eqref{eq:bla} using the $X$ measurements and separately again using the calibrated $Y$ measurements (since they are de-biased). We refer to these estimates as $\hat{s}_x$ and $\hat{s}_y$, which we then aggregate via a weighted average and define $\hat{s}_i = (r_{x,i}\hat{s}_{x,i} + r_{y,i}\hat{s}_{y,i})/(r_{x,i}+r_{y,i})$, $i=1,2,\ldots,n$. 

To construct confidence intervals for the conditional PoA we again use bootstrapping to avoid reliance on restrictive distributional assumptions. Whereas \citet{taffe2023use} uses a parametric bootstrap that assumes $s$ is normally distributed, our proposed bootstrap method is fully nonparametric. In particular, we repeat the conditional PoA calculation described above on each of $B$ bootstrap resamples yielding $\left\{\widehat{\text{PoA}}(\hat{s}_i)^*_1, \widehat{\text{PoA}}(\hat{s}_i)^*_2, \ldots, \widehat{\text{PoA}}(\hat{s}_i)^*_B\right\}$. Importantly, to construct confidence intervals that reflect sampling variation for a given subject $i$, each bootstrap sample is taken with replacement conditional on the guaranteed inclusion of subject $i$. For the sake of numerical stability, and because the CIs of interest are pointwise for each subject, we simply use the percentile method; confidence limits are given by the $\alpha/2$ and $1-\alpha/2$ percentiles of the $B$ bootstrap estimates $\left\{\widehat{\text{PoA}}(\hat{s}_i)^*_1, \widehat{\text{PoA}}(\hat{s}_i)^*_2, \ldots, \widehat{\text{PoA}}(\hat{s}_i)^*_B\right\}$. We demonstrate the application of this methodology in Section \ref{sec:ex}.

\section{Example: Comparison of tPSA Measurement Methods} \label{sec:ex}

Here we illustrate the proposed probability of agreement analysis in the context of a method comparison study undertaken to compare methods of measuring total Prostatic Specific Antigen (tPSA, $\mu$g/L) measurements. In particular, we analyze the results from \citet{ferraro2023managing}, which compares the Roche Cobas e801 with three other analytical platforms: Abbot Alinity i, Beckman Access Dxl, and Siemens Atellica IM. The study involved $n=135$ subjects whose tPSA values were measured $r=2$ times by all four of the methods. For illustration, we specifically compare the Roche and Siemens methods, treating them respectively as the reference and comparator.

\begin{figure}[htbp]
    \centering
    \includegraphics[width=0.8\textwidth]{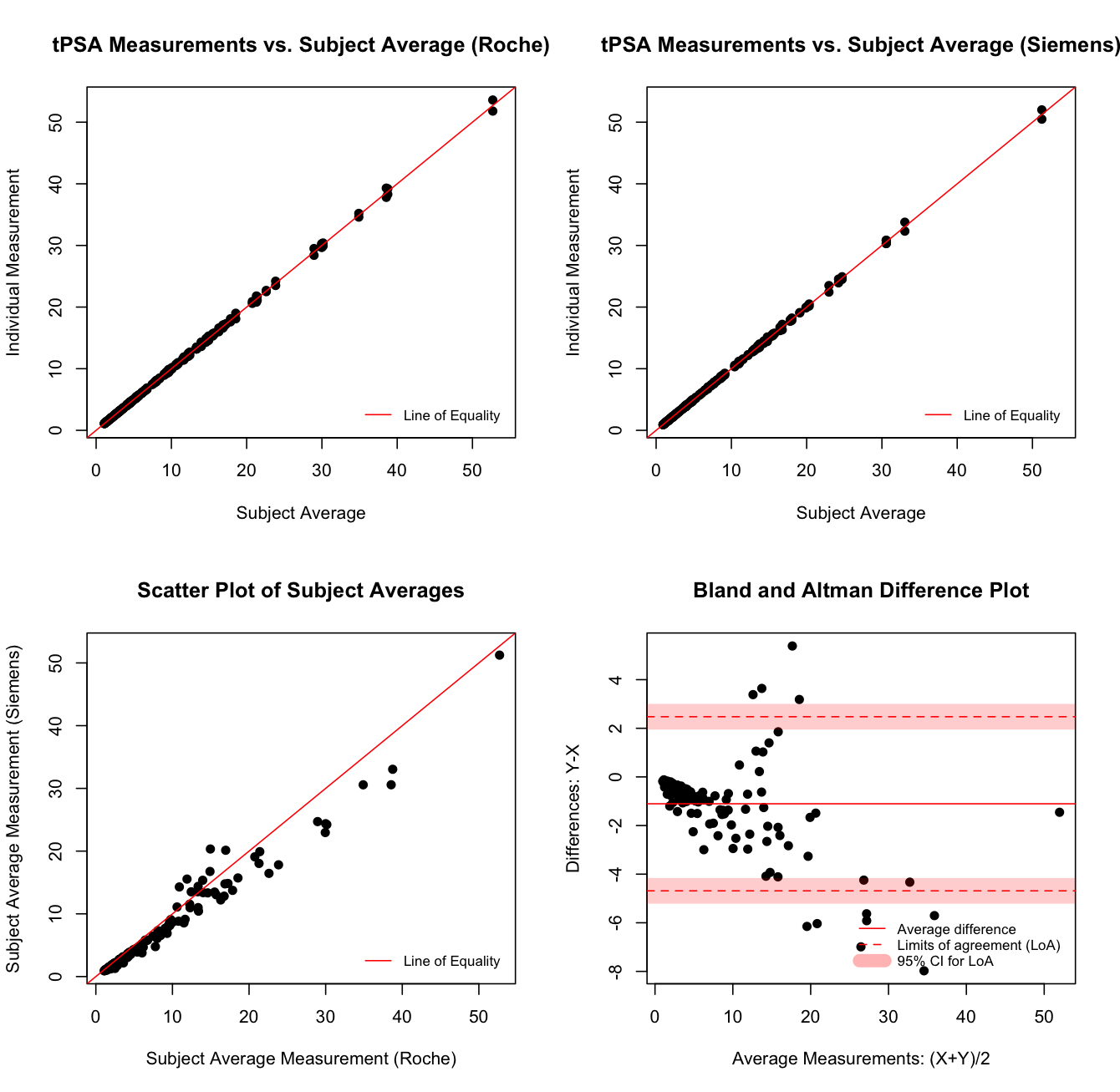}
    \caption{Visual summaries of agreement between methods Roche ($X$) and Siemens ($Y$). Top Left: Scatter plot of tPSA measurements by Roche vs. the average tPSA per subject. Top Right: Scatter plot of tPSA measurements by Siemens vs. the average tPSA per subject. Bottom Left: Scatter plot of Siemens subject-averages versus Roche subject-averages. Bottom Right: Bland-Altman plot of subject-averages.}
    \label{fig:data_viz}
\end{figure}

We begin by visually summarizing the data. The scatter plots in the top panels of Figure \ref{fig:data_viz} visualize the individual measurements taken by the Roche (left) and Siemens (right) methods, with the tPSA replicates for each subject plotted against the average of those replicates. Overall, we see very consistent tPSA measurements within each method (the points are clustered along the lines of equality in both plots) suggesting high precision. The bottom left panel plots the subject-averages by both methods against each other. Because we see a clear deviation in these points away from the line of equality, this indicates that a relative bias exists between the two methods, with the Siemens method generally producing smaller measurements and the magnitude of this discrepancy increasing with tPSA values. Additionally, this scatterplot suggests that measurement variability increases as the true underlying tPSA values increase since the dispersion of points increases for larger measured values. These informal insights suggest that a model like \eqref{eq:model} is needed to capture the apparent bias and non-constant measurement variation. The bottom right panel displays the Bland-Altman plot for these data, which visualizes the agreement between methods by plotting the difference in method-specific subject-averages versus the overall average measurement on each subject. This plot corroborates the insights described previously: the differences deviate substantially from zero and the nature of this deviation changes with tPSA (indicating a non-constant relative bias) and the dispersion of the differences increases with tPSA (indicating non-constant measurement variability). 

\begin{table}[]
\centering
\begin{tabular}{|ccc|}
\hline
Parameter & Estimate & 95\% Confidence Interval           \\ \hline
$\mu$        & 8.6414   & (7.1367, 10.1460)  \\
$\sigma$     & 8.8956   & (6.8693, 10.9219)  \\
$\beta_0$     & --0.2103  & (--0.3575, --0.0630) \\
$\beta_1$     & 0.9009   & (0.8537, 0.9482)   \\
$\omega_{x,0}$   & --0.0022  & (--0.0255, 0.0212)  \\
$\omega_{x,1}$   & 0.0194   & (0.0155, 0.0232)   \\
$\omega_{y,0}$   & 0.0288   & (0.0001, 0.0576)   \\
$\omega_{y,1}$   & 0.0058   & (--0.0027, 0.0142)  \\
$\omega_{y,2}$   & 0.0002   & (--0.0001, 0.0006)  \\ \hline
\end{tabular}
\caption{Parameter estimates and confidence intervals for the tPSA example.}
\label{tab:param_est}
\end{table}

\begin{figure}[htbp]
    \centering
    \includegraphics[width=1\textwidth]{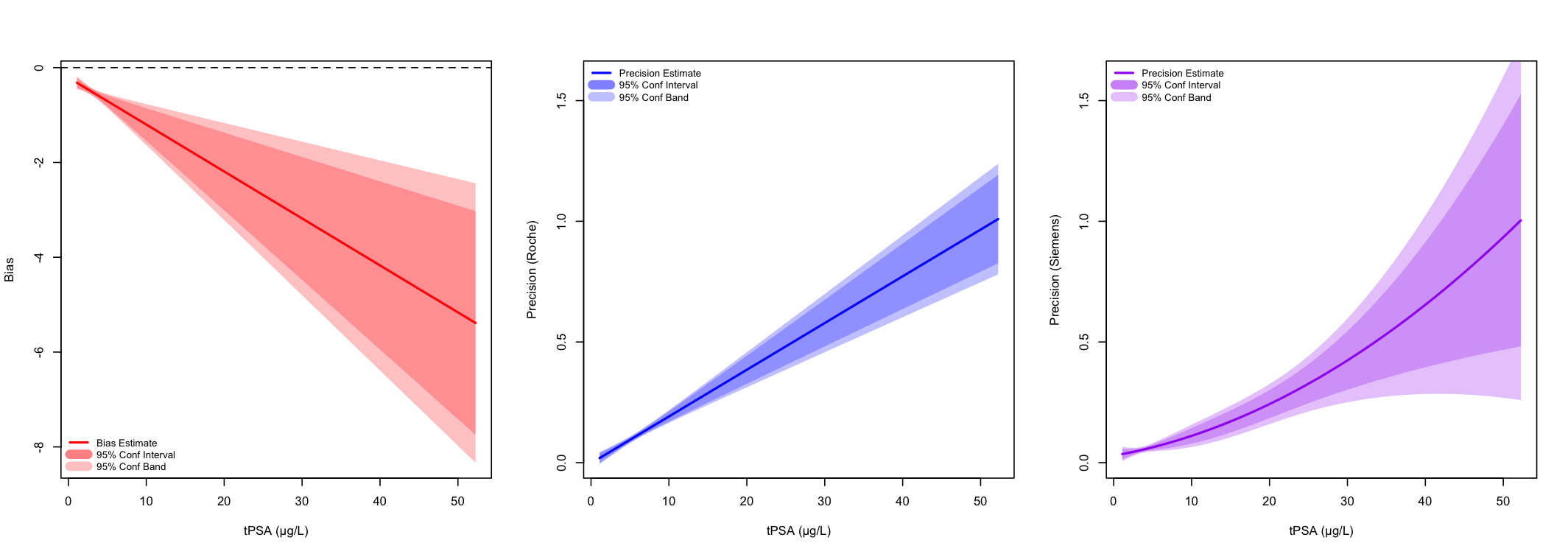}
    \caption{Visualizations of bias (left), reference method precision (middle), comparator method precision (left).}
    \label{fig:bias_prec}
\end{figure}

We formalize this assessment of agreement with the probability of agreement analysis. We begin by fitting model \eqref{eq:model} for a variety of different orders $(p,d_x,d_y)$ and find via $BIC$ selection that $(p=1,d_x=1,d_y=2)$ is a good choice. The estimates and 95\% confidence intervals for the parameters of this model are given in Table \ref{tab:param_est}. Note the confidence intervals are \textit{standard} bootstrap intervals with $B=10,000$. These estimates and CIs provide further evidence of significant relative bias and heterscedasticity. With these estimates, we can estimate (and visualize) the relative bias and heteroscedaccity of these methods; Figure \ref{fig:bias_prec} visualizes (what \citet{taffe2019methodcompare} refers to as) bias and precision plots. We see a clear relative bias, whereby Siemens measurements tend to be lower than Roche measurements, and increasingly so as tPSA increases. We also see that the measurement variation by each method increases with tPSA, though the nature of this relationship is not the same for each method. In all cases, there is increased uncertainty for large tPSA values, which is due to the small number of observed tPSA readings in this range. Note that these plots visualize both the 95\% pointwise confidence intervals as well as the 95\% simultaneous confidence bands using the standard bootstrap methods described in Section \ref{sec:ci_stand}\footnote{
Versions of the bias, precision, and PoA plots with uncertainty visualized by \textit{percentile-based} confidence intervals/bands are provided in Appendix \hyperref[sec:appxC]{C}. The code used to produce these results are available on GitHub.}.

\begin{figure}[htbp]
    \centering
    \includegraphics[width=1\textwidth]{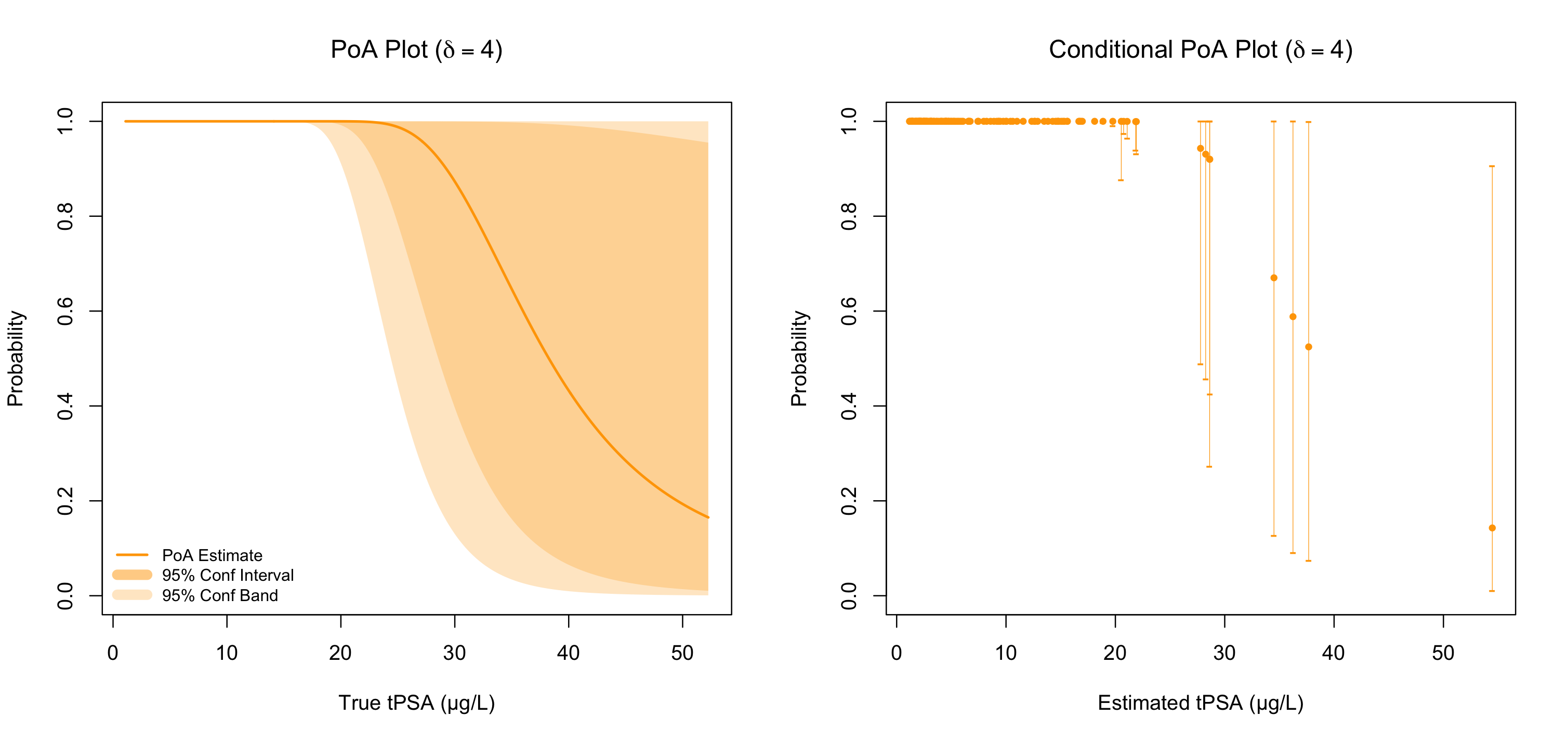}
    \includegraphics[width=1\textwidth]{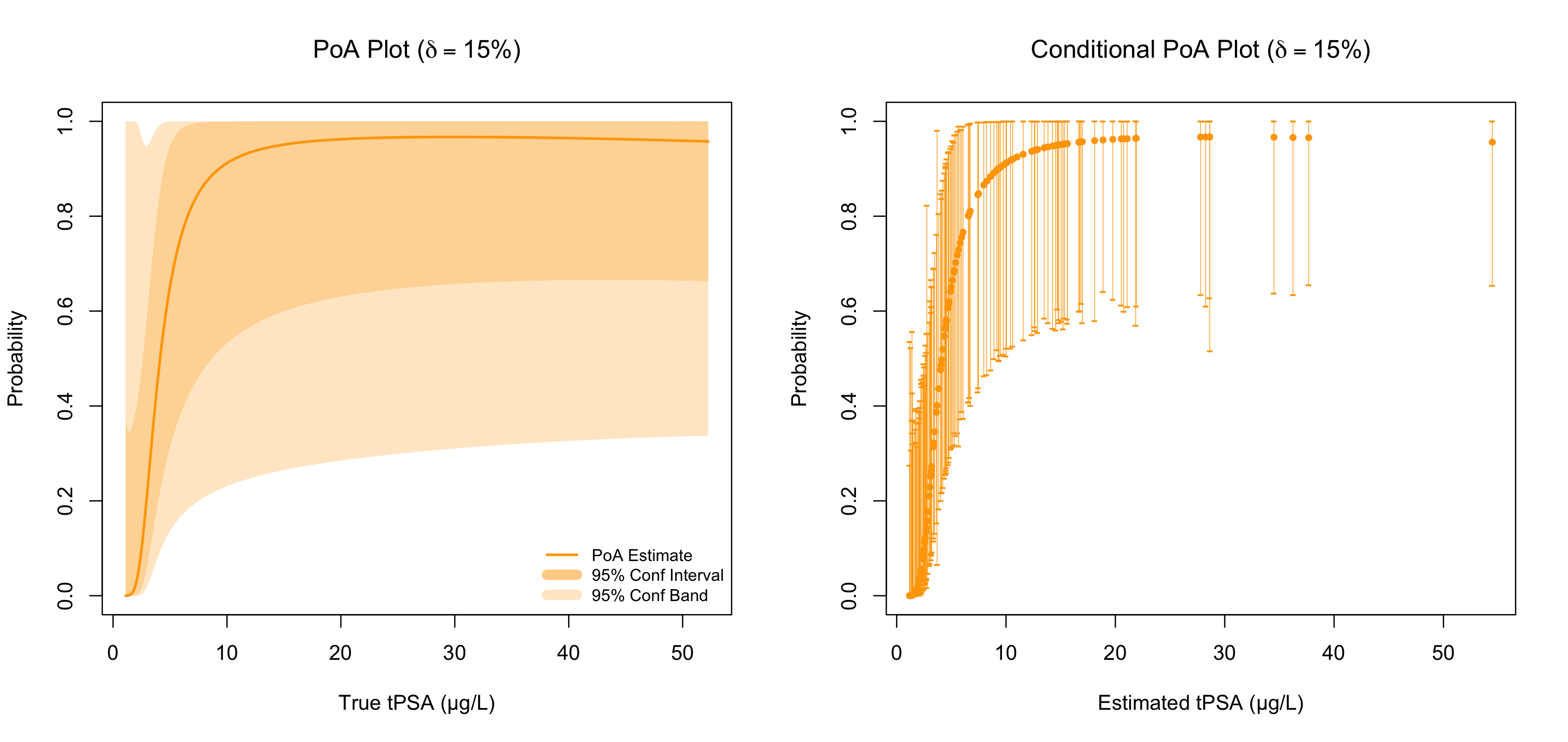}
    \caption{Probability of agreement (left) and conditional probability of agreement (right) with a fixed (top) and relative (bottom) equivalence margin.}
    \label{fig:poa}
\end{figure}

All of the analyses so far indicate moderate agreement between the Roche and Siemens methods for smaller tPSA values and a decline in agreement as tPSA increases. We formally explore this insight with the probability of agreement. To perform this analysis we require a choice for $\delta$. For illustration, we choose $\delta=4$, implying that tPSA readings within $\pm4$ $\mu$g/L of each other are clinically equivalent. The top left panel of Figure \ref{fig:poa} depicts the corresponding PoA plot which indicates that it is very likely that Roche and Siemens measurements are clinically equivalent if the true underlying tPSA value is less than 20 $\mu$g/L, but agreement sharply declines for larger tPSA values. This assessment is applicable to a population of individuals. If we wish to draw analogous conclusions for the specific individuals in the study, we consult the conditional probability of agreement plot (top right panel of Figure \ref{fig:poa}). We draw broadly similar conclusions in this case. 

This assessment of agreement assumed that the equivalence margin $\delta$ was fixed. However, in certain circumstances, we may be able to tolerate larger differences for larger values of the true underlying trait. For instance, suppose two measurements are deemed clinically equivalent if their difference is within 15\% of the true latent trait. The bottom panels of Figure \ref{fig:poa} display plots of the probability of agreement (left) and conditional probability of agreement (right), constructed with $\delta(s) = 0.15s$. We see that with a relative equivalence margin, the pattern of agreement fundamentally changes; it is quite unlikely to expect tPSA measurements within $\pm 15\%$ of each other for small tPSA values, but quite likely that this tolerance is met for large tPSA values. These conclusions may be applied to the specific individuals in this study as well as the broader population. 

\begin{figure}[htbp]
    \centering
    \includegraphics[width=1\textwidth]{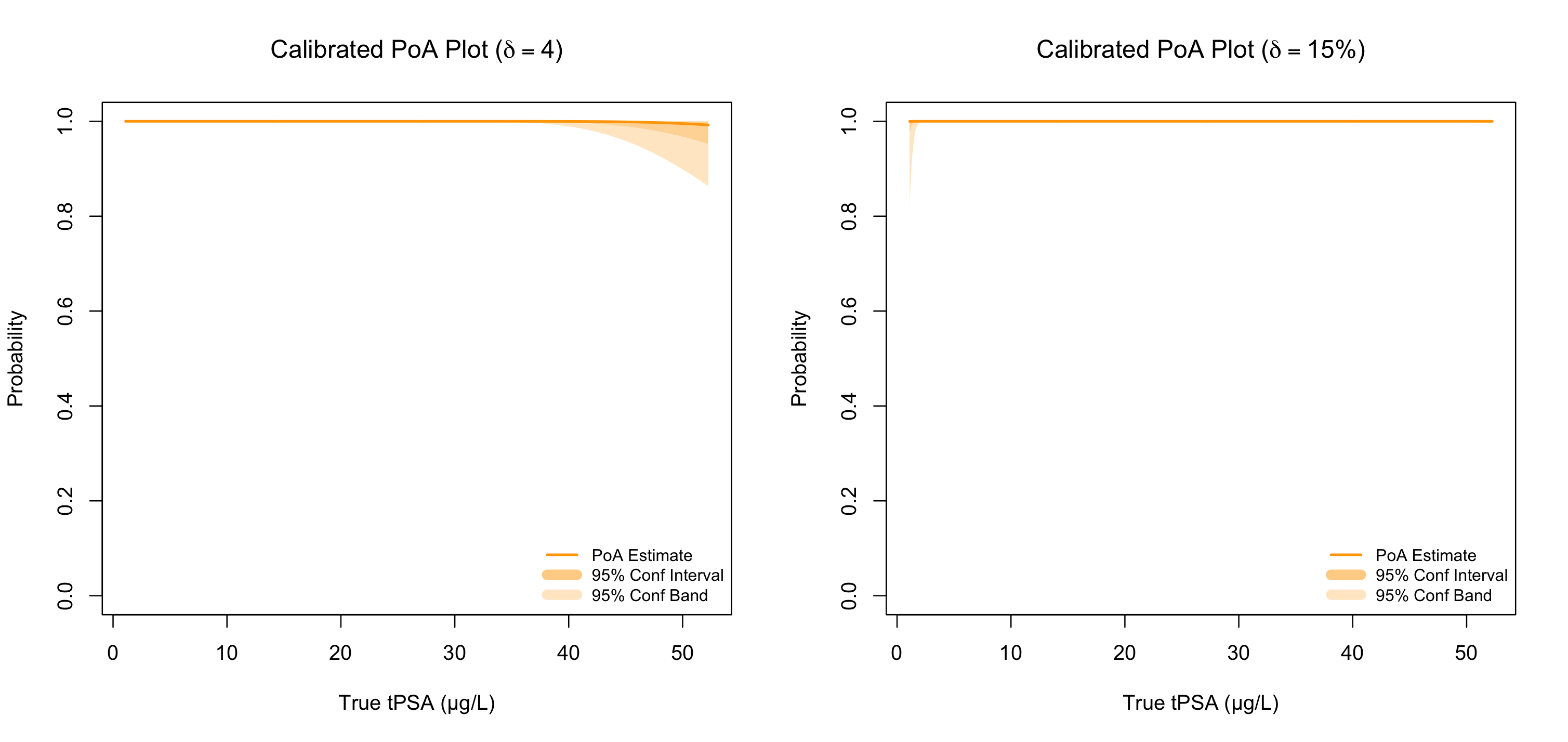}
    \caption{Calibrated probability of agreement with fixed (left) or relative (right) equivalence margin.}
    \label{fig:cal_poa}
\end{figure}

Regardless of how we specified $\delta$, we concluded that there are some tPSA values for which the Roche and Siemens methods do not agree. As discussed above, relative bias is one component of this disagreement. By appropriately calibrating the Siemens measurements, this source of disagreement can be remedied. Figure \ref{fig:cal_poa} displays the probability of agreement \textit{after} calibration for both choices of $\delta$. In both cases, we now see much better agreement across the range of plausible tPSA values. Thus, it might not be appropriate to use the Roche and Siemens methods interchangeably for all subjects, but interchanging the Roche method with a calibrated version of the Siemens method may be permissible. 

\section{Numerical Studies} \label{sec:sim}

Having illustrated the proposed methodology in a specific example, we now study its performance more broadly. In particular, and across a wide variety of scenarios, we examine the quality of the estimates produced as well as the coverage of the associated confidence intervals (Section \ref{sec:sim_inf}), and we also investigate the accuracy of the proposed $BIC$-based order selection procedure (Section \ref{sec:sim_bic}). We begin, in Section \ref{sec:sim_scen}, by describing the design of the aforementioned numerical investigations.

\subsection{Simulation Design} \label{sec:sim_scen}

To investigate the performance of the proposed methodology, we repeatedly generate data from model \eqref{eq:model} with $n=100$, $S\sim \text{UNIF}(10,40)$\footnote{Note: the entire simulation was repeated with $S\sim \text{N}(25,75)$ and $S\sim \text{GAM}(25/3,3)$, but the results were nearly indistinguishable from the $S\sim \text{UNIF}(10,40)$ case and therefore not presented here.}, and a particular specification of the parameters. Each time, an appropriate model is selected, the relevant parameters are estimated, and the 95\% confidence intervals/bands are constructed. For the parameter settings, we consider the eight scenarios detailed in Table \ref{tab:sim_scen} which span various magnitudes and structures for bias and precision. Note that Scenarios 1-5 are the same as those considered by \citet{taffe2020assessing,taffe2023use}, and Scenarios 6-8 specify more complicated bias and precision structures.

For each parameter scenario, we also consider three replication scenarios (see Table \ref{tab:sim_scen}) which differ in the number of replicate measurements made by each method. The full simulation considers all 24 combinations of the parameter-replication scenarios, and in each scenario we simulate data 1000 times and then calculate the bias and root mean squared error (RMSE) associated with the parameter estimates, as well as the coverage of the associated confidence intervals. For the PoA, we calculate bias and RMSE averaged over the values of $s$ considered, and \textit{simultaneous} coverage is calculated for the confidence bands. The confidence intervals/bands investigated here are the standard bootstrap ones described in Section \ref{sec:ci_stand}. Performance of the percentile-based bootstrap intervals/bands is expected to be at least as good, given the parameter-replication settings considered. 

For each parameter-replicate scenario, we also investigate the performance of the proposed $BIC$-based order selection scheme. For this, we simulate each of the 24 scenarios 100 times and report the proportion of those in which the polynomial orders were correctly chosen. Note that this investigation is computationally taxing, which explains the use of 100 not 1000 iterations like the inference investigation. 

\begin{table}[]
\centering
\begin{tabular}{|cc|}
\hline
\multicolumn{2}{|c|}{Parameter Scenarios}   \\ \hline
\multicolumn{1}{|c|}{1}         & $\boldsymbol{\beta}=(0,1)^\top$, $\boldsymbol{\omega}_x = (1.75, 0.08)^\top$, $\boldsymbol{\omega}_y = (0, 0.2)^\top$   \\
\multicolumn{1}{|c|}{2}         & $\boldsymbol{\beta}=(-6,0.85)^\top$, $\boldsymbol{\omega}_x = (0.15, 0.09)^\top$, $\boldsymbol{\omega}_y = (0.1, 0.07)^\top$ \\
\multicolumn{1}{|c|}{3}         & $\boldsymbol{\beta}=(-4,1.2)^\top$, $\boldsymbol{\omega}_x = (2, 0.01)^\top$, $\boldsymbol{\omega}_y = (1, 0.05)^\top$         \\
\multicolumn{1}{|c|}{4}         & $\boldsymbol{\beta}=(4,0.8)^\top$, $\boldsymbol{\omega}_x = (1.75, 0.08)^\top$, $\boldsymbol{\omega}_y = (0, 0.2)^\top$         \\
\multicolumn{1}{|c|}{5}         & $\boldsymbol{\beta}=(4,1.2)^\top$, $\boldsymbol{\omega}_x = (1.75, 0.08)^\top$, $\boldsymbol{\omega}_y = (0, 0.2)^\top$         \\
\multicolumn{1}{|c|}{6}         & $\boldsymbol{\beta}=(8.3,-0.4, 0.03)^\top$, ${\omega}_x = 3$, ${\omega}_y = 4$         \\
\multicolumn{1}{|c|}{7}         & $\boldsymbol{\beta}=(-8.3,2.4, -0.03)^\top$, $\boldsymbol{\omega}_x = (2, 0.01)^\top$, $\boldsymbol{\omega}_y = (1, 0.05)^\top$         \\
\multicolumn{1}{|c|}{8}         & $\boldsymbol{\beta}=(20.3125,-2.9375, 0.1875, -0.0025)^\top$, $\boldsymbol{\omega}_x = (5.4, -1, 0.06, -0.0008)^\top$, $\boldsymbol{\omega}_y = (5.44, -0.98, 0.062, -0.00083)^\top$         \\ \hline
\multicolumn{2}{|c|}{Replication Scenarios} \\ \hline
\multicolumn{1}{|c|}{1}         & $r_{x,i}\sim\text{UNIF}(2,5)$, $r_{y,i}\sim\text{UNIF}(2,5)$ \\
\multicolumn{1}{|c|}{2}         & $r_{x,i}\sim\text{UNIF}(9,11)$, $r_{y,i}\sim\text{UNIF}(2,5)$ \\
\multicolumn{1}{|c|}{3}         & $r_{x,i}\sim\text{UNIF}(9,11)$, $r_{y,i}\sim\text{UNIF}(9,11)$ \\ \hline
\end{tabular}
    \caption{Simulation Scenarios.}
    \label{tab:sim_scen}
\end{table}

\subsection{Simulation Results: Inference} \label{sec:sim_inf}

\begin{figure}[]
    \centering
    \includegraphics[width=1\textwidth]{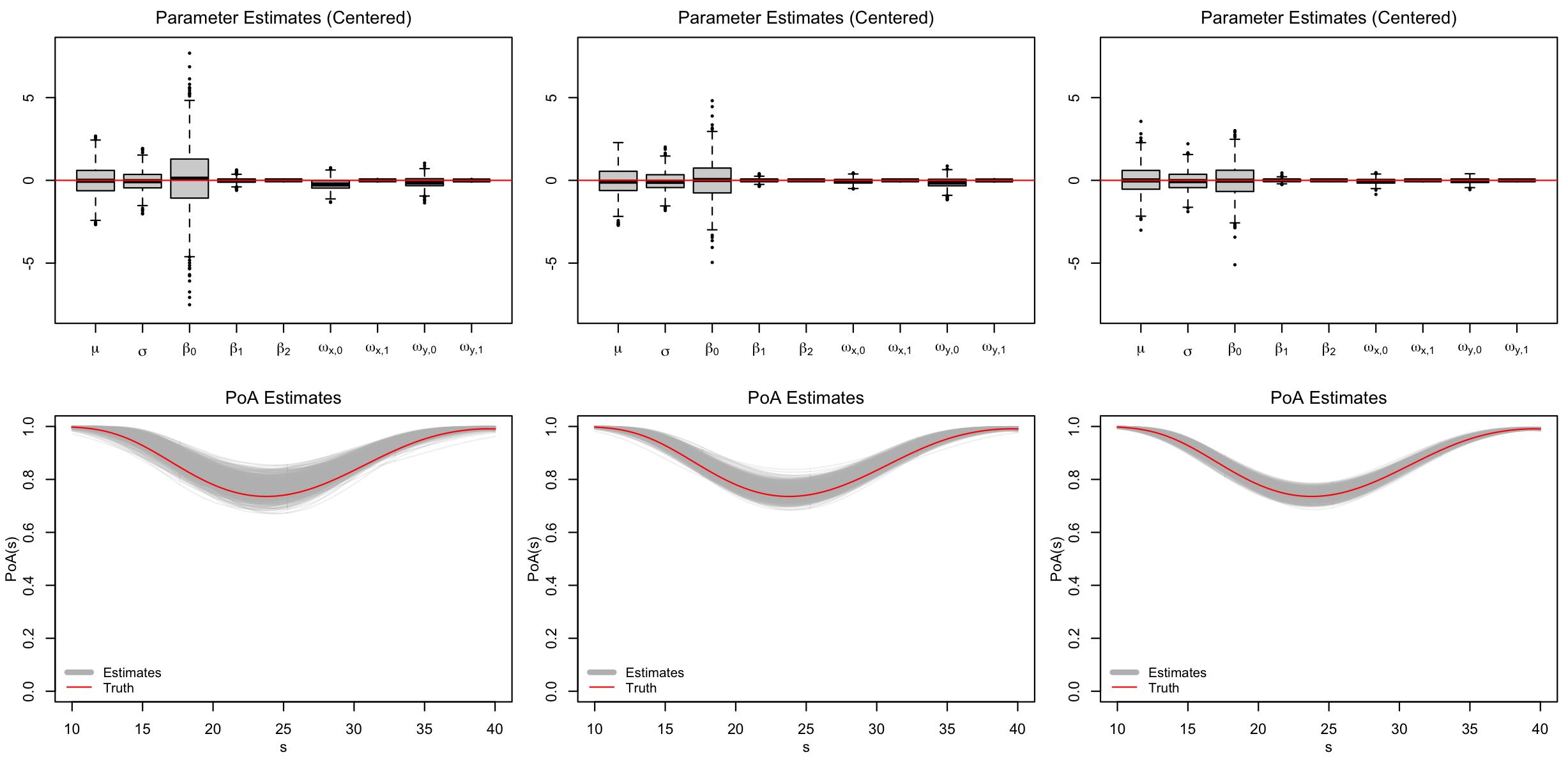}
    \caption{Simulation results for Parameter Scenario 7, displayed for Replication Scenarios 1 (left column), 2 (middle column), and 3 (right columns).}
    \label{fig:sim_scen7}
\end{figure}

The results for each parameter scenario (across the three replication scenarios) are summarized in both graphical and tabular format. We present the results for parameter scenario 7 here, and include the remaining scenarios' results in Appendix \hyperref[sec:appxD]{D}. The figures include, for each parameter, boxplots of the 1000 estimates (centered by the parameters' true values), as well as a visualization of the 1000 PoA estimates with the true PoA curve overlaid. The tables include the bias, RMSE, and coverage values for each parameter in each replication scenario. From Figure \ref{fig:sim_scen7} and Table \ref{tab:sim_scen7} we draw the following insights, which apply broadly across all scenarios investigated. For both the individual model parameters as well as the PoA more generally, bias and RMSE are acceptably low, and both improve as the number of observations in the study increase. Likewise, coverage tends to be close to nominal, and especially so for larger study sizes. We find that increasing the number of replicates for a given method improves the estimation of that method's precision parameters, and generally speaking, we see that estimation uncertainty tends to increase as the complexity of the bias and precision polynomials increases. Even still, with a moderate number of subjects, and a small amount of replication by each method, one can expect the estimated PoA curve to yield trustworthy insights, with the level of trust increasing with the size of the comparison study.

\begin{table}[]
\centering
\begin{tabular}{|l|c|c|c|}
\hline
Parameter & Bias & RMSE & Coverage \\ \hline
$\mu=25$ & 0.002 / 0.026 / -0.004 & 0.942 / 0.879 / 0.906 & 0.936 / 0.946 / 0.937 \\ 
$\sigma=8.66$ & -0.005 / 0.005 / 0.005 & 0.437 / 0.393 / 0.394 & 0.946 / 0.952 / 0.962 \\ 
$\beta_0=-8.3$ & 1.570 / 0.498 / 0.472 & 2.610 / 1.462 / 1.181 & 0.850 / 0.912 / 0.910 \\ 
$\beta_1=2.4$ & -0.145 / -0.046 / -0.045 & 0.228 / 0.129 / 0.102 & 0.844 / 0.919 / 0.907 \\ 
$\beta_2=-0.03$ & 0.003 / 0.001 / 0.001 & 0.004 / 0.003 / 0.002 & 0.853 / 0.916 / 0.903 \\ 
$\omega_{x,0}=2$ & -0.230 / -0.049 / -0.062 & 0.387 / 0.163 / 0.168 & 0.868 / 0.945 / 0.923 \\ 
$\omega_{x,1}=0.01$ & -0.001 / -0.000 / -0.000 & 0.012 / 0.006 / 0.006 & 0.939 / 0.945 / 0.940 \\ 
$\omega_{y,0}=1$ & -0.103 / -0.109 / -0.033 & 0.316 / 0.310 / 0.152 & 0.935 / 0.920 / 0.945 \\ 
$\omega_{y,1}=0.05$ & -0.006 / -0.006 / -0.001 & 0.014 / 0.014 / 0.006 & 0.911 / 0.917 / 0.938 \\ 
$\text{PoA}$ & 0.028 / 0.015 / 0.007 &  0.034 / 0.020 / 0.013 & 0.908 / 0.956 / 0.983 \\ \hline
\end{tabular}
\caption{Simulation results for Parameter Scenario 7, displayed for Replication Scenarios 1/2/3.}
\label{tab:sim_scen7}
\end{table}

\subsection{Simulation Results: Order Selection} \label{sec:sim_bic}

Table \ref{tab:sim_bic} displays the proportion of simulation runs in which the $BIC$-based order selection method (described in Section \ref{sec:model}) selects the correct order. Broadly speaking, we see excellent performance: order selection for the bias polynomial $g(s;\boldsymbol{\beta})$ has near-perfect accuracy with even limited replication, and although order selection is slightly more difficult for the precision polynomials $\sigma_x(s;\boldsymbol{\omega}_x)$ and $\sigma_y(s;\boldsymbol{\omega}_y)$, accuracy is also near-perfect with increased replication. That said, we see poorer performance in parameter scenario 8, where the orders of all polynomials are reasonably high ($p=d_x=d_y=3$). To investigate this, we define Replication Scenario 4 ($r_{x,i}, r_{y,i}\sim\text{UNIF}(20,25)$) and find that with enough replication, the $BIC$-based selection method does indeed work as expected. But this suggests that in cases with very complicated bias and precision structures, one should consider increasing replication and/or augmenting the order selection process with information from other sources such as residual diagnostics or significance tests.

\begin{table}[]
\centering
\begin{tabular}{|c|c|c|c|}
\hline
    Parameter Scenario       & $g(s;\boldsymbol{\beta})$ & $\sigma_x(s;\boldsymbol{\omega}_x)$ & $\sigma_y(s;\boldsymbol{\omega}_y)$ \\ \hline
1 & 0.98 / 1.00 / 1.00 / 1.00 & 0.76 / 0.99 / 0.95 / 1.00 & 0.96 / 0.99 / 0.93 / 1.00 \\ 
2 & 0.93 / 0.99 / 1.00 / 1.00 & 0.90 / 0.93 / 0.95 / 1.00 & 0.97 / 1.00 / 0.98 / 1.00 \\ 
3 & 0.99 / 1.00 / 1.00 / 1.00 & 1.00 / 0.99 / 1.00 / 1.00 & 1.00 / 0.85 / 0.97 / 0.99 \\ 
4 & 1.00 / 1.00 / 1.00 / 1.00 & 0.73 / 1.00 / 0.96 / 1.00 & 0.96 / 0.97 / 0.97 / 1.00 \\ 
5 & 0.98 / 0.98 / 1.00 / 1.00 & 0.76 / 0.97 / 0.97 / 1.00 & 0.96 / 1.00 / 0.96 / 1.00 \\ 
6 & 0.93 / 0.99 / 1.00 / 1.00 & 1.00 / 0.97 / 1.00 / 1.00 & 0.99 / 0.99 / 0.97 / 1.00 \\ 
7 & 0.95 / 1.00 / 1.00 / 1.00 & 0.99 / 1.00 / 0.99 / 1.00 & 0.88 / 0.70 / 0.96 / 0.99 \\ 
8 & 0.40 / 0.68 / 0.78 / 1.00 & 0.42 / 0.62 / 0.70 / 0.98 & 0.41 / 0.60 / 0.71 / 0.98 \\ \hline
\end{tabular}
\caption{$BIC$-based order selection results for each parameter scenario and polynomial, displayed for replication scenarios 1/2/3/4.}
\label{tab:sim_bic}
\end{table}

\section{Conclusion} \label{sec:conc}

In this paper we extend the probability of agreement (PoA) methodology \citep{stevens2017assessing, stevens2018comparing, taffe2023use} so that it may be more flexibly applied to the analysis of method comparison studies. In particular, we develop an inference framework, and provide accompanying code, that (i) does not require a distributional assumption for the underlying characteristic being measured, (ii) accommodates possibly non-linear bias and precision, (iii) accommodates flexibility in the specification of clinical equivalence, and (iv) allows for imbalance in the number of replicate measurements by each method on each subject. These contributions make an already intuitive and effective methodology even more accessible and practically useful. 

However, there exist other helpful extensions and opportunities for further refinement. One such extension would be to generalize model \eqref{eq:model} to a more complicated linear mixed-effect model that captures other effects and sources of variation. For instance, if the measurement methods were operated by multiple individuals, or if the subjects could be grouped into homogeneous subgroups, it would be important to account for this. See \citet{parker2020using} for examples of such models and contexts. Another extremely valuable extension would be the development of a sample size determination method for this analysis. While we explored the effect of sample size in this paper, it would be helpful to determine--\textit{ahead} of the comparison study--how many subjects and replicate measurements are needed to ensure the probability of agreement analysis is trustworthy. \citet{hagar2023precision} have explored this for a simpler version of the probability of agreement, determining subject and replicate numbers that ensure the PoA is estimated with sufficient precision, so this seems like a promising direction for future work.

\appendix
\section*{Appendices} \label{sec:appx}
\setcounter{table}{0}
\renewcommand{\thetable}{A\arabic{table}}
\setcounter{figure}{0}
\renewcommand{\thefigure}{A\arabic{figure}}

\subsection*{Appendix A: Best Linear Approximation} \label{sec:appxA}
The best linear approximation of $S_i$ based on $X_{ij}$ is $\gamma_0+\gamma_1X_{ij}$ where $\gamma_0$ and $\gamma_1$ are chosen to minimize $\text{E}\left[(S_i - \gamma_0-\gamma_1X_{ij})^2\right]$. It is straightforward to show that we must have $\gamma_0=\text{E}[S_{i}]-\gamma_1\text{E}[X_{ij}]$ and $\gamma_1=\text{Cov}[S_i, X_{ij}]/\text{Var}[X_{ij}]$. Thus, the best linear approximation of $S_i$ based on $X_{ij}$ is \citep{carroll2006measurement} $$\text{E}[S_i]+\frac{\text{Cov}[S_i,{X}_{ij}]}{\text{Var}[{X}_{ij}]}\left({X}_{ij}-\text{E}[X_{ij}]\right).$$ Because we have replicate measurements on subject $i$, this can be extended to the best linear approximation of $S_i$ based on $\overline{X}_i$ \citep{carroll2006measurement}: $$\text{E}[S_i]+\frac{\text{Cov}[S_i,\overline{X}_i]}{\text{Var}[\overline{X}_i]}\left(\overline{X}_i-\text{E}[\overline{X}_{i}]\right),$$ which is then estimated by $$\hat\mu+\frac{\hat{\sigma}^2}{\hat\sigma^2+\hat\sigma^2_{x,i}/r_{x,i}}\left(\overline{x}_i-\hat\mu\right),$$ since $\text{E}[S_i]=\text{E}[X_{ij}]=\text{E}[\overline{X}_{i}]=\mu$, $\text{Cov}[S_i, \overline{X}_i]=\text{Var}[S_i]=\sigma^2$ and $\text{Var}[\overline{X}_i]=\sigma^2+\sigma^2_{x,i}/r_{x,i}$.

\newpage
\noindent \textbf{Proof that} $\boldsymbol{\textbf{Cov}[S_i, \overline{X}_i]=\textbf{Var}[S_i]}$
\begin{align*}
\text{Cov}\left[S_i, \overline{X}_i\right] &= \text{Cov}\left[S_i, \frac{1}{r_{x,i}}\sum_{j=1}^{r_{x,i}}X_{ij}\right] \\ 
&= \text{Cov}\left[S_i, \frac{1}{r_{x,i}}\sum_{j=1}^{r_{x,i}}(S_{i}+M_{ij})\right] \\
&= \text{Cov}[S_i, S_i] + \text{Cov}\left[S_i, \frac{1}{r_{x,i}}\sum_{j=1}^{r_{x,i}}M_{ij}\right] \\ 
&= \text{Var}[S_i]+\frac{1}{r_{x,i}}\sum_{j=1}^{r_{x,i}}\text{Cov}[S_i, M_{ij}] \\
&= \sigma^2
\end{align*}
since $\text{Cov}[S_i, M_{ij}]=\text{E}[S_iM_{ij}]=0$ for all $j=1,2,\ldots,r_{x,i}$ by the law of total expectation. \qed

\noindent \textbf{Proof that} $\boldsymbol{\textbf{Var}[\overline{X}_i]=\sigma^2+\sigma^2_{x,i}/r_{x,i}}$
\begin{align*}
\text{Var}[\overline{X}_i]&=\text{Var}\left[\frac{1}{r_{x,i}}\sum_{j=1}^{r_{x,i}}X_{ij}\right] \\ 
&= \text{Var}\left[\frac{1}{r_{x,i}}\sum_{j=1}^{r_{x,i}}(S_i+M_{ij})\right] \\
&= \text{Var}\left[S_i+\frac{1}{r_{x,i}}\sum_{j=1}^{r_{x,i}}M_{ij}\right] \\ 
&= \text{Var}[S_i]+\frac{1}{r_{x,i}^2}\sum_{j=1}^{r_{x,i}}\text{Var}[M_{ij}] + 2\text{Cov}\left[S_i,\frac{1}{r_{x,i}}\sum_{j=1}^{r_{x,i}}M_{ij}\right]\\ 
&= \sigma^2+\frac{\sigma^2_{x,i}}{r_{x,i}}+\frac{2}{r_{x,i}}\sum_{j=1}^{r_{x,i}}\text{Cov}[S_i, M_{ij}] \\ 
&= \sigma^2+\frac{\sigma^2_{x,i}}{r_{x,i}}
\end{align*}
since $\text{Cov}[S_i, M_{ij}]=0$ for all $j=1,2,\ldots,r_{x,i}$ as above. \qed

The estimates $\hat\mu$, $\hat\sigma^2$, and $\hat\sigma^2_{x,i}$ are found by the method of moments, exploiting the following expectations.
\begin{align*}
\text{E}[\overline{X}] = \text{E}\left[\frac{1}{N}\sum_{i=1}^n\sum_{j=1}^{r_{x,i}}X_{ij}\right] = \frac{1}{N}\sum_{i=1}^n\sum_{j=1}^{r_{x,i}}\text{E}\left[X_{ij}\right] =\frac{1}{N}\sum_{i=1}^n\sum_{j=1}^{r_{x,i}}\mu = \mu
\end{align*}
\begin{align*}
\text{E}\left[\sum_{j=1}^{r_{x,i}}\left(X_{ij}-\overline{X}_i\right)^2\right] &= \text{E}\left[\sum_{j=1}^{r_{x,i}}\left(\left(X_{ij}-\mu\right)-\left(\overline{X}_i-\mu\right)\right)^2\right] \\
&= \sum_{j=1}^{r_{x,i}}\text{E}\left[\left(X_{ij}-\mu\right)^2\right] + \sum_{j=1}^{r_{x,i}}\text{E}\left[\left(\overline{X}_{i}-\mu\right)^2\right] -2\sum_{j=1}^{r_{x,i}}\text{E}\left[\left(X_{ij}-\mu\right)\left(\overline{X}_{i}-\mu\right)\right]\\
&= \sum_{j=1}^{r_{x,i}}\text{Var}\left[X_{ij}\right] + \sum_{j=1}^{r_{x,i}}\text{Var}\left[\overline{X}_{i}\right] -2\sum_{j=1}^{r_{x,i}}\text{Cov}\left[X_{ij},\overline{X}_{i}\right]\\
&= \sum_{j=1}^{r_{x,i}}(\sigma^2+\sigma^2_{x,i}) + \sum_{j=1}^{r_{x,i}}\left(\sigma^2+\frac{\sigma^2_{x,i}}{r_{x,i}}\right) -2\sum_{j=1}^{r_{x,i}}\left(\sigma^2+\frac{\sigma^2_{x,i}}{r_{x,i}}\right)\\
&=(r_{x,i}-1)~\sigma^2_{x,i}
\end{align*}
\begin{align*}
\text{E}\left[\sum_{i=1}^{n}\sum_{j=1}^{r_{x,i}}\left(\overline{X}_i-\overline{X}\right)^2\right] &= \text{E}\left[\sum_{i=1}^n\sum_{j=1}^{r_{x,i}}\left(\left(\overline{X}_i-\mu\right)-\left(\overline{X}-\mu\right)\right)^2\right] \\
&= \sum_{i=1}^{n}\sum_{j=1}^{r_{x,i}}\text{E}\left[\left(\overline{X}_{i}-\mu\right)^2\right] + \sum_{i=1}^{n}\sum_{j=1}^{r_{x,i}}\text{E}\left[\left(\overline{X}-\mu\right)^2\right] -2\sum_{i=1}^{n}\sum_{j=1}^{r_{x,i}}\text{E}\left[\left(\overline{X}_{i}-\mu\right)\left(\overline{X}-\mu\right)\right]\\
&= \sum_{i=1}^{n}\sum_{j=1}^{r_{x,i}}\text{Var}\left[\overline{X}_{i}\right] + \sum_{i=1}^{n}\sum_{j=1}^{r_{x,i}}\text{Var}\left[\overline{X}\right] -2\sum_{i=1}^{n}\sum_{j=1}^{r_{x,i}}\text{Cov}\left[\overline{X}_{i},\overline{X}\right]\\
&= \sum_{i=1}^{n}\sum_{j=1}^{r_{x,i}}\left(\sigma^2+\frac{\sigma^2_{x,i}}{r_{x,i}}\right) + \sum_{i=1}^{n}\sum_{j=1}^{r_{x,i}}\left\{\frac{\sigma^2}{N^2}\sum_{i=1}^nr_{x,i}^2+\frac{1}{N^2}\sum_{i=1}^nr_{x,i}\sigma^2_{x,i}\right\} -2\sum_{i=1}^{n}\sum_{j=1}^{r_{x,i}}\left(\frac{r_{x,i}\sigma^2 + \sigma^2_{x,i}}{N}\right)\\
&= \sum_{i=1}^{n}\left(r_{x,i}\sigma^2+\sigma^2_{x,i}\right) + \frac{\sigma^2}{N}\sum_{i=1}^nr_{x,i}^2+\frac{1}{N}\sum_{i=1}^nr_{x,i}\sigma^2_{x,i}-\frac{2}{N}\sum_{i=1}^{n}\left(r_{x,i}^2\sigma^2 + r_{x,i}\sigma^2_{x,i}\right)\\
&=\sigma^2\left(N-\frac{1}{N}\sum_{i=1}^{n}r_{x,i}^2\right) + \sum_{i=1}^{n}\left(1-\frac{r_{x,i}}{N}\right)\sigma^2_{x,i}
\end{align*}
Simultaneously solving 
\begin{align*}
\text{E}[\overline{X}] &= \mu \\
\text{E}\left[\sum_{j=1}^{r_{x,i}}\left(X_{ij}-\overline{X}_i\right)^2\right] &= (r_{x,i}-1)~\sigma^2_{x,i} \\
\text{E}\left[\sum_{i=1}^{n}\sum_{j=1}^{r_{x,i}}\left(\overline{X}_i-\overline{X}\right)^2\right] &= \sigma^2\left(N-\frac{1}{N}\sum_{i=1}^{n}r_{x,i}^2\right) + \sum_{i=1}^{n}\left(1-\frac{r_{x,i}}{N}\right)\sigma^2_{x,i}
\end{align*}
for each parameter and substituting the observed data yields the following method of moments estimates:
\begin{align*}
\hat\mu &= \frac{1}{N}\sum_{i=1}^n\sum_{j=1}^{r_{x,i}}x_{ij} \\
\hat\sigma^2_{x,i} &= \frac{1}{r_{x,i}-1}\sum_{j=1}^{r_{x,i}}(x_{ij}-\overline{x}_i)^2 \\
\hat\sigma^2 &= \frac{\sum_{i=1}^{n}r_{x,i}\left(\overline{x}_i-\overline{x}\right)^2-\sum_{i=1}^n(1-r_{x,i}/N)\hat\sigma^2_{x,i}}{N-\frac{1}{N}\sum_{i=1}^nr^2_{x,i}}
\end{align*}


\newpage

\subsection*{Appendix B: Partial Derivatives for the Delta Method Confidence Bands} \label{sec:appxB}

Recall that the probability of agreement is defined as $$\text{PoA}(s)=\Phi\left(\frac{\delta_U(s)-g(s;\boldsymbol{\beta})+s}{\sqrt{\sigma_x^2(s;\boldsymbol{\omega}_x) + \sigma_y^2(s;\boldsymbol{\omega}_y)}}\right) - \Phi\left(\frac{\delta_L(s)-g(s;\boldsymbol{\beta})+s}{\sqrt{\sigma_x^2(s;\boldsymbol{\omega}_x) + \sigma_y^2(s;\boldsymbol{\omega}_y)}}\right).$$ Note that this generalizes Equation \eqref{eq:poa2}, taking the interval of clinically acceptable differences to be $(\delta_L(s),\delta_U(s))$ instead of $(-\delta,\delta)$. Since the PoA must be contained within [0,1], we define $\eta(s;\boldsymbol{\theta})=f(\text{PoA}(s))$ where $\boldsymbol{\theta}=(\mu,\sigma,\boldsymbol{\beta}^\top, \boldsymbol{\omega}_x^\top, \boldsymbol{\omega}_y^\top)^\top$ and $f(p)=\log(-\log(1-p))$ is the complementary log-log function. As described in Section \ref{sec:ci_stand}, we require the partial derivative vector $\frac{\partial\eta(s;\boldsymbol{\theta})}{\partial\boldsymbol{\theta}}$. By the chain rule we have: 
\begin{align*}
\frac{\partial\eta(s;\boldsymbol{\theta})}{\partial\boldsymbol{\theta}} &= \left(\frac{\partial\eta(s;\boldsymbol{\theta})}{\partial\mu},\frac{\partial\eta(s;\boldsymbol{\theta})}{\partial\sigma},\frac{\partial\eta(s;\boldsymbol{\theta})}{\partial\boldsymbol{\beta}}^\top,\frac{\partial\eta(s;\boldsymbol{\theta})}{\partial\boldsymbol{\omega}_x}^\top,\frac{\partial\eta(s;\boldsymbol{\theta})}{\partial\boldsymbol{\omega}_y}^\top\right)^\top \\ &= f^\prime(\text{PoA}(s))\left(\frac{\partial\text{PoA}(s)}{\partial\mu},\frac{\partial\text{PoA}(s)}{\partial\sigma},\frac{\partial\text{PoA}(s)}{\partial\boldsymbol{\beta}}^\top,\frac{\partial\text{PoA}(s)}{\partial\boldsymbol{\omega}_x}^\top,\frac{\partial\text{PoA}(s)}{\partial\boldsymbol{\omega}_y}^\top\right)^\top
\end{align*}
where $f^\prime(p)=\frac{1}{(p-1)\log(1-p)}$, $\frac{\partial\text{PoA}(s)}{\partial\mu}=\frac{\partial\text{PoA}(s)}{\partial\sigma}=0$ and
$$\frac{\partial\text{PoA}(s)}{\partial\boldsymbol{\beta}} = \left[\phi\left(\frac{\delta_L(s)-g(s;\boldsymbol{\beta})+s}{\sqrt{\sigma_x^2(s;\boldsymbol{\omega}_x) + \sigma_y^2(s;\boldsymbol{\omega}_y)}}\right)-\phi\left(\frac{\delta_U(s)-g(s;\boldsymbol{\beta})+s}{\sqrt{\sigma_x^2(s;\boldsymbol{\omega}_x) + \sigma_y^2(s;\boldsymbol{\omega}_y)}}\right)\right]\times\frac{\left(1, s, s^2, \ldots, s^p\right)^\top}{\left(\sigma_x^2(s;\boldsymbol{\omega}_x) + \sigma_y^2(s;\boldsymbol{\omega}_y)\right)^{1/2}}$$

\begin{align*}
\frac{\partial\text{PoA}(s)}{\partial\boldsymbol{\omega}_x} = &\left[\phi\left(\frac{\delta_L(s)-g(s;\boldsymbol{\beta})+s}{\sqrt{\sigma_x^2(s;\boldsymbol{\omega}_x) + \sigma_y^2(s;\boldsymbol{\omega}_y)}}\right)(\delta_L(s)-g(s;\boldsymbol{\beta})+s) -\phi\left(\frac{\delta_U(s)-g(s;\boldsymbol{\beta})+s}{\sqrt{\sigma_x^2(s;\boldsymbol{\omega}_x) + \sigma_y^2(s;\boldsymbol{\omega}_y)}}\right)(\delta_U(s)-g(s;\boldsymbol{\beta})+s)\right] \\ &\times\frac{\sigma_x(s;\boldsymbol{\omega}_x)\left(1, s, s^2, \ldots, s^{d_x}\right)^\top}{\left(\sigma_x^2(s;\boldsymbol{\omega}_x) + \sigma_y^2(s;\boldsymbol{\omega}_y)\right)^{3/2}}
\end{align*}

\begin{align*}
\frac{\partial\text{PoA}(s)}{\partial\boldsymbol{\omega}_y} = &\left[\phi\left(\frac{\delta_L(s)-g(s;\boldsymbol{\beta})+s}{\sqrt{\sigma_x^2(s;\boldsymbol{\omega}_x) + \sigma_y^2(s;\boldsymbol{\omega}_y)}}\right)(\delta_L(s)-g(s;\boldsymbol{\beta})+s) -\phi\left(\frac{\delta_U(s)-g(s;\boldsymbol{\beta})+s}{\sqrt{\sigma_x^2(s;\boldsymbol{\omega}_x) + \sigma_y^2(s;\boldsymbol{\omega}_y)}}\right)(\delta_U(s)-g(s;\boldsymbol{\beta})+s)\right] \\ &\times\frac{\sigma_y(s;\boldsymbol{\omega}_y)\left(1, s, s^2, \ldots, s^{d_y}\right)^\top}{\left(\sigma_x^2(s;\boldsymbol{\omega}_x) + \sigma_y^2(s;\boldsymbol{\omega}_y)\right)^{3/2}}
\end{align*}

\noindent Note that $\phi(\cdot)$ above is the standard normal probability density function.

In the case of constructing confidence bands for bias, define $\text{Bias}(s) = \beta_0+\beta_{1}s +\cdots+\beta_{p}s^p-s \equiv \eta(s;\boldsymbol{\beta})$, in which case we have: $\frac{\partial\eta(s;\boldsymbol{\beta})}{\partial\boldsymbol{\beta}} = \left(1, s, s^2, \ldots, s^p\right)^\top$. And in the case of constructing confidence bands for the precision functions, define $\sigma(s;\boldsymbol{\omega}) = \omega_{0}+\omega_{1}s +\cdots+\omega_{d}s^{d} \equiv \eta(s;\boldsymbol{\omega})$, in which case we have: $\frac{\partial\eta(s;\boldsymbol{\omega})}{\partial\boldsymbol{\omega}} = \left(1, s, s^2, \ldots, s^d\right)^\top$.

\newpage

\subsection*{Appendix C: Additional tPSA Example Results} \label{sec:appxC}

\begin{table}[!h]
\centering
\begin{tabular}{c|ccccc|}
\cline{2-6}
\multicolumn{1}{l|}{}      & \multicolumn{5}{c|}{Order}                                \\ \cline{2-6} 
                           & 0         & 1         & 2         & 3         & 4         \\ \hline
\multicolumn{1}{|c|}{$g(s;\boldsymbol{\beta})$}    & ---        & 216.4738  & 223.7031  & 331.4890  & 219.4773  \\
\multicolumn{1}{|c|}{$\sigma_x(s;\boldsymbol{\omega}_x)$} & --427.6165 & --610.6658 & --608.8373 & --606.6254 & --596.9571 \\
\multicolumn{1}{|c|}{$\sigma_y(s;\boldsymbol{\omega}_y)$} & --467.8196 & --581.7699 & --588.2871 & --582.9635 & --576.8539 \\ \hline
\end{tabular}
\caption{$BIC$ values for polynomials of different orders.}
\label{tab:bic}
\end{table}

\begin{figure}[!h]
    \centering
    \includegraphics[width=0.75\textwidth]{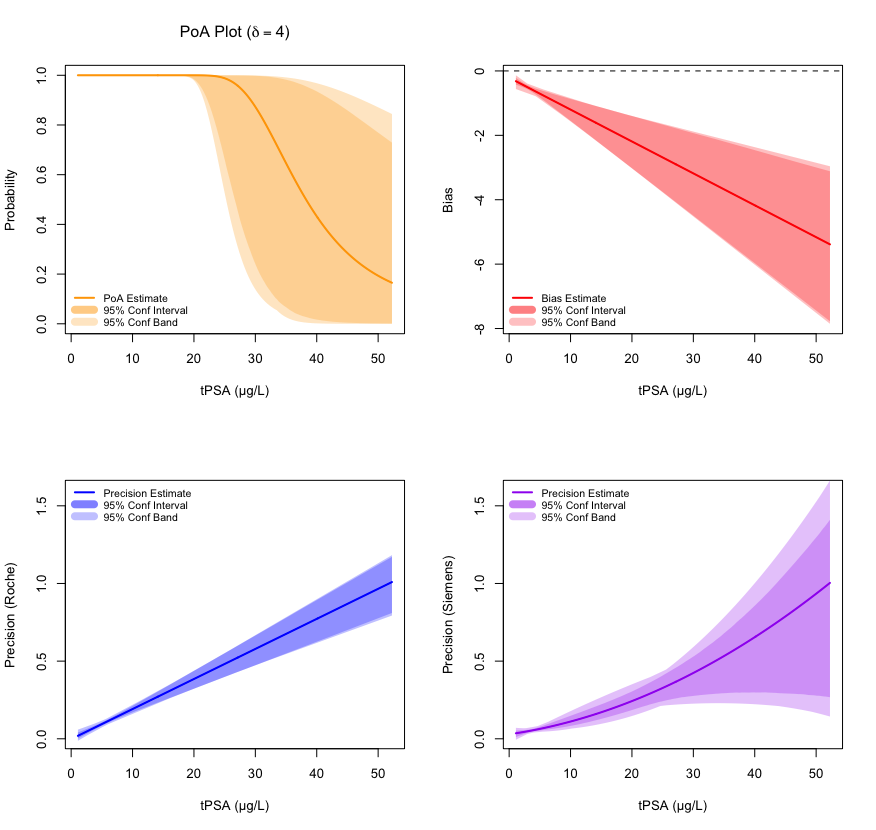}
    \caption{PoA (top left), bias (top right), and precision (bottom) plots for tPSA data with confidence intervals/bans constructed via percentile bootstrap methods.}
    \label{fig:bias_prec_cal}
\end{figure}


\newpage

\subsection*{Appendix D: Additional Simulation Results} \label{sec:appxD}

\begin{figure}[H]
    \centering
    \includegraphics[width=1\textwidth]{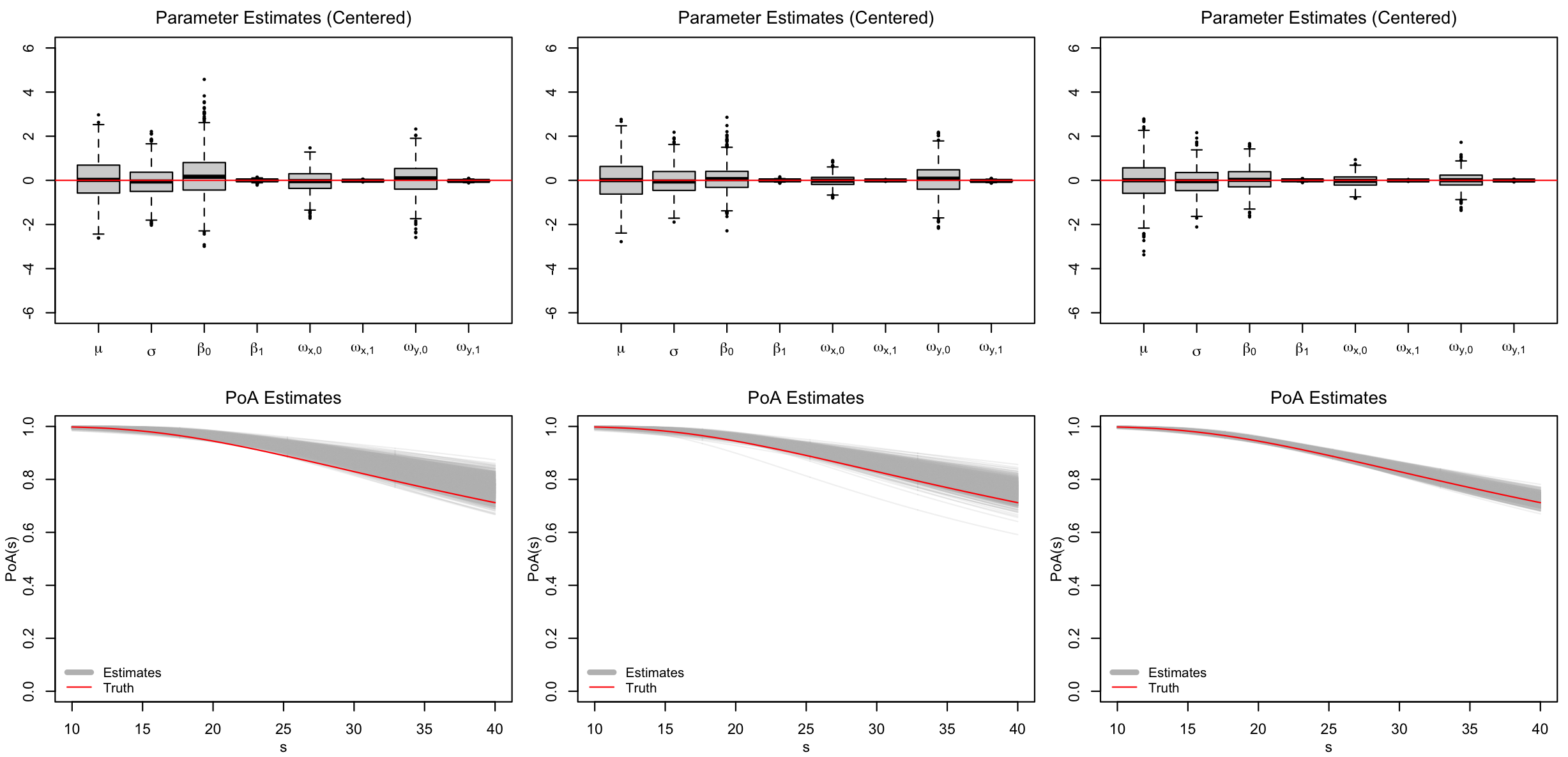}
    \caption{Simulation results for Parameter Scenario 1, displayed for Replication Scenarios 1 (left column), 2 (middle column), and 3 (right columns).}
    \label{fig:sim_scen1}
\end{figure}

\begin{table}[H]
\centering
\begin{tabular}{|l|c|c|c|}
\hline
Parameter & \multicolumn{1}{c|}{Bias} &  \multicolumn{1}{c|}{RMSE} & \multicolumn{1}{c|}{Coverage} \\ \hline
$\mu=25$ & 0.001 / 0.031 / -0.041 & 0.966 / 0.864 / 0.895 & 0.926 / 0.951 / 0.937 \\
$\sigma=8.66$ & -0.024 / -0.007 / -0.013 & 0.444 / 0.405 / 0.408 & 0.949 / 0.957 / 0.952 \\
$\beta_0=0$ & 0.496 / 0.118 / 0.083 &  1.084 / 0.738 / 0.496 & 0.919 / 0.940 / 0.933 \\
$\beta_1=1$ & -0.016 / -0.003 / -0.001 & 0.047 / 0.035 / 0.023 & 0.937 / 0.948 / 0.939 \\
$\omega_{x,0}=1.75$ & -0.076 / -0.036 / -0.034 & 0.478 / 0.251 / 0.242 & 0.942 / 0.947 / 0.963 \\
$\omega_{x,1}=0.08$ & -0.012 / -0.003 / -0.003 & 0.024 / 0.011 / 0.011 & 0.882 / 0.934 / 0.949 \\
$\omega_{y,0}=0$ & 0.069 / -0.037 / 0.019 & 0.651 / 0.659 / 0.333 & 0.942 / 0.944 / 0.944 \\
$\omega_{y,1}=0.2$ & -0.025 / -0.022 / -0.006 & 0.039 / 0.038 / 0.017 & 0.840 / 0.867 / 0.909 \\
$\text{PoA}$ & 0.034 / 0.024 / 0.008 & 0.037 / 0.027 / 0.011 & 0.700 / 0.807 / 0.968 \\ \hline
\end{tabular}
\caption{Simulation results for Parameter Scenario 1, displayed for Replication Scenarios 1/2/3.}
\label{tab:sim_scen1}
\end{table}

\begin{figure}[H]
    \centering
    \includegraphics[width=1\textwidth]{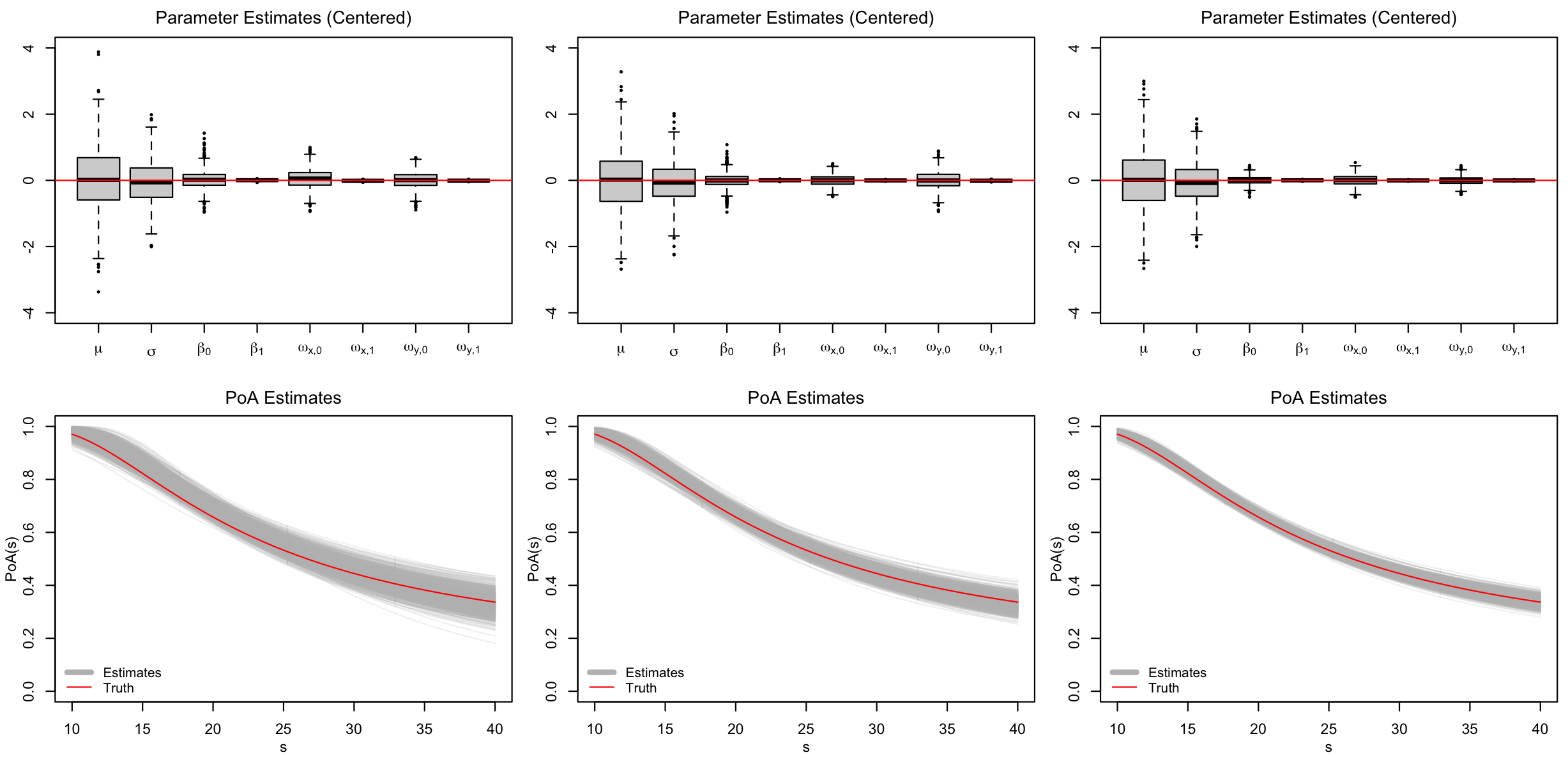}
    \caption{Simulation results for Parameter Scenario 2, displayed for Replication Scenarios 1 (left column), 2 (middle column), and 3 (right columns).}
    \label{fig:sim_scen2}
\end{figure}

\begin{table}[H]
\centering
\begin{tabular}{|l|c|c|c|}
\hline
Parameter & \multicolumn{1}{c|}{Bias} & \multicolumn{1}{c|}{RMSE} & \multicolumn{1}{c|}{Coverage} \\ \hline
$\mu=25$ & -0.005 / 0.002 / -0.004 & 0.919 / 0.879 / 0.842 & 0.946 / 0.949 / 0.950 \\
$\sigma=8.66$ & -0.040 / -0.008 / -0.008 & 0.460 / 0.388 / 0.389 & 0.936 / 0.957 / 0.957 \\
$\beta_0=-6$ & 0.041 / 0.006 / 0.004 & 0.373 / 0.271 / 0.197 & 0.953 / 0.939 / 0.942 \\
$\beta_1=0.85$ & 0.003 / 0.001 / 0.001 & 0.018 / 0.013 / 0.010 & 0.947 / 0.943 / 0.946 \\
$\omega_{x,0}=0.15$ & 0.071 / 0.007 / 0.006 & 0.320 / 0.157 / 0.156 & 0.938 / 0.944 / 0.943 \\
$\omega_{x,1}=0.09$ & -0.014 / -0.003 / -0.003 & 0.020 / 0.008 / 0.008 & 0.805 / 0.921 / 0.922 \\
$\omega_{y,0}=0.1$ & -0.003 / -0.003 / -0.001  & 0.246 / 0.246 / 0.117 & 0.941 / 0.943 / 0.949 \\
$\omega_{y,1}=0.07$ & -0.008 / -0.009 / -0.002 & 0.014 / 0.015 / 0.006 & 0.866 / 0.845 / 0.932 \\
$\text{PoA}$ & 0.016 / 0.006 / 0.005 & 0.035 / 0.022 / 0.015 & 0.968 / 0.990 / 0.989 \\ \hline
\end{tabular}
\caption{Simulation results for Parameter Scenario 2, displayed for Replication Scenarios 1/2/3.}
\label{tab:sim_scen2}
\end{table}

\begin{figure}[H]
    \centering
    \includegraphics[width=1\textwidth]{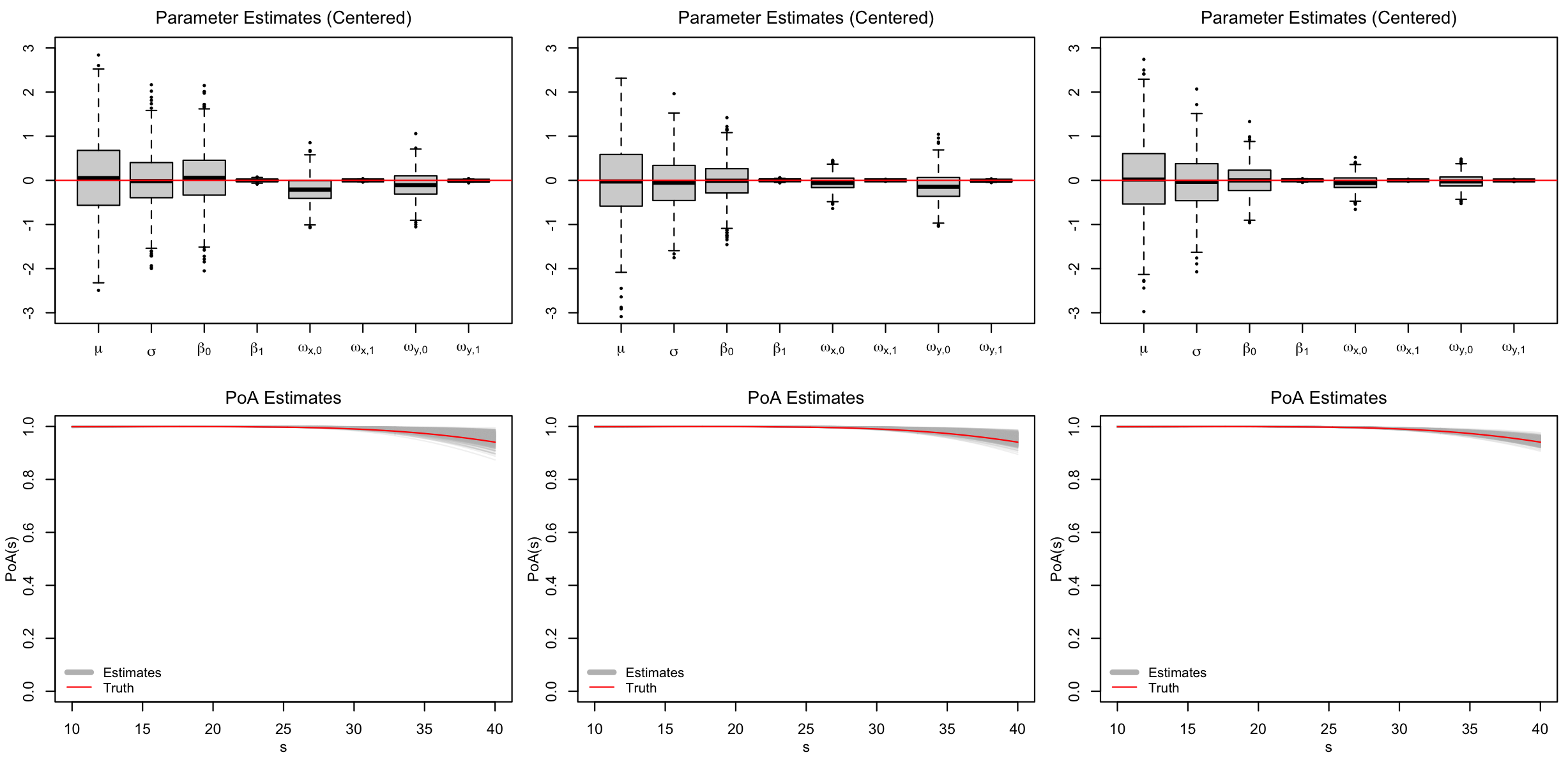}
    \caption{Simulation results for Parameter Scenario 3, displayed for Replication Scenarios 1 (left column), 2 (middle column), and 3 (right columns).}
    \label{fig:sim_scen3}
\end{figure}

\begin{table}[H]
\centering
\begin{tabular}{|l|c|c|c|}
\hline
Parameter & \multicolumn{1}{c|}{Bias}  & \multicolumn{1}{c|}{RMSE} & \multicolumn{1}{c|}{Coverage} \\ \hline
$\mu=25$ & 0.041 / 0.011 / 0.019 & 0.911 / 0.867 / 0.915 & 0.946 / 0.949 / 0.934 \\
$\sigma=8.66$ & 0.004 / 0.003 / 0.000 & 0.441 / 0.387 / 0.409 & 0.942 / 0.960 / 0.939 \\
$\beta_0=-4$ & 0.119 / 0.028 / -0.005 & 0.634 / 0.437 / 0.340 & 0.943 / 0.933 / 0.929 \\
$\beta_1=1.2$ & -0.005 / -0.001 / 0.000 & 0.025 / 0.018 / 0.013 & 0.942 / 0.933 / 0.933 \\
$\omega_{x,0}=2$ & -0.236 / -0.063 / -0.050 & 0.396 / 0.168 / 0.165 & 0.875 / 0.929 / 0.933 \\
$\omega_{x,1}=0.01$ & -0.002 / 0.000 / 0.000 & 0.012 / 0.006 / 0.006 & 0.939 / 0.949 / 0.947 \\
$\omega_{y,0}=1$ & -0.133 / -0.090 / -0.032  & 0.324 / 0.302 / 0.153 & 0.923 / 0.926 / 0.930 \\
$\omega_{y,1}=0.05$ & -0.006 / -0.006 / -0.001 & 0.014 / 0.014 / 0.007 & 0.935 / 0.911 / 0.933 \\
$\text{PoA}$ & 0.005 / 0.003 / 0.001 & 0.005 / 0.004 / 0.002 & 0.648 / 0.784 / 0.944 \\ \hline
\end{tabular}
\caption{Simulation results for Parameter Scenario 3, displayed for Replication Scenarios 1/2/3.}
\label{tab:sim_scen3}
\end{table}

\begin{figure}[H]
    \centering
    \includegraphics[width=1\textwidth]{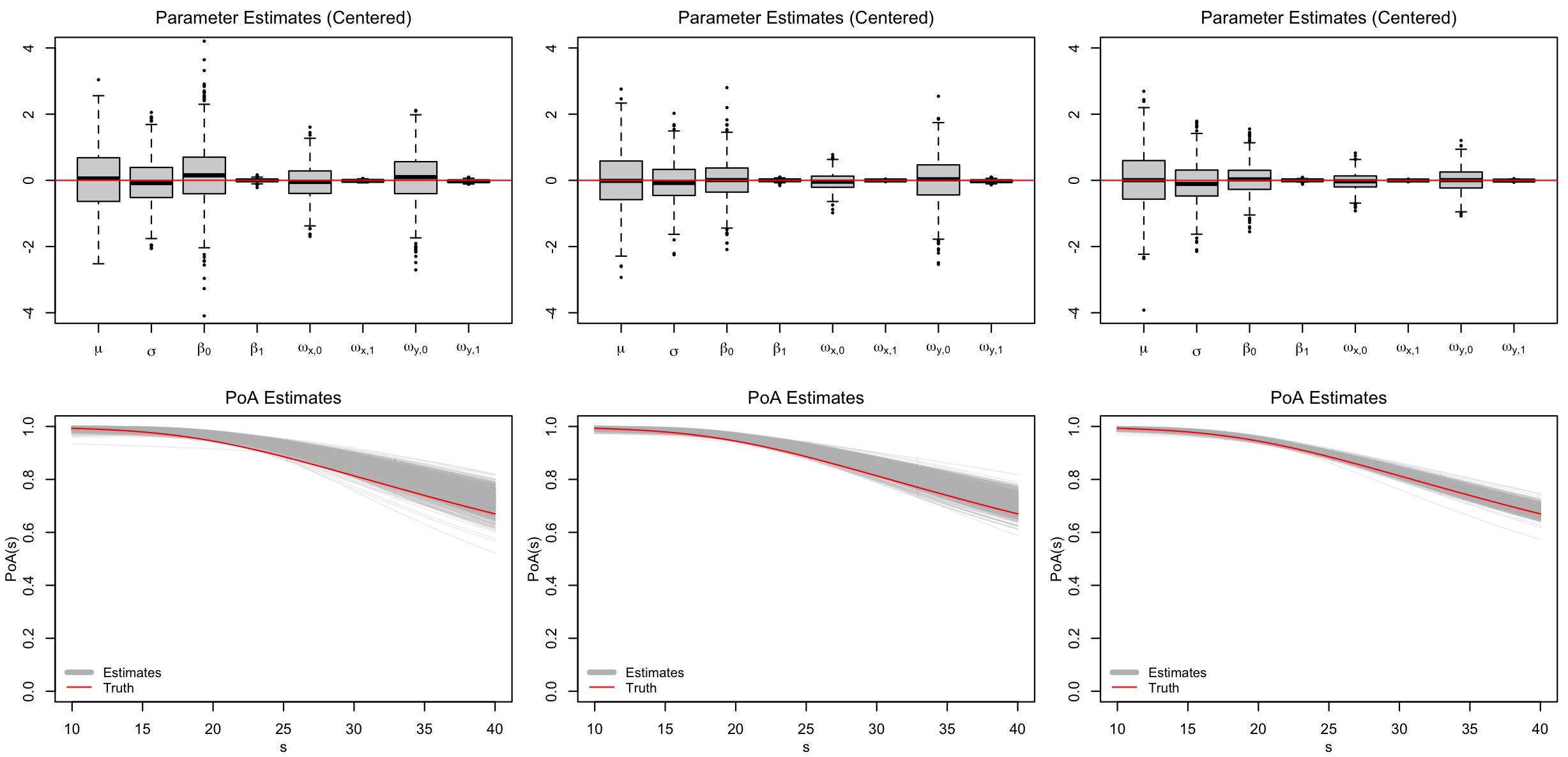}
    \caption{Simulation results for Parameter Scenario 4, displayed for Replication Scenarios 1 (left column), 2 (middle column), and 3 (right columns).}
    \label{fig:sim_scen4}
\end{figure}

\begin{table}[H]
\centering
\begin{tabular}{|l|c|c|c|}
\hline
Parameter & Bias & RMSE & Coverage \\ \hline
$\mu=25$ & 0.040 / -0.001 / 0.050 & 0.935 / 0.909 / 0.882 & 0.941 / 0.948 / 0.943 \\ 
$\sigma=8.66$ & -0.024 / -0.013 / -0.001 & 0.475 / 0.430 / 0.423 & 0.937 / 0.945 / 0.949 \\ 
$\beta_0=4$ & 0.356 / 0.076 / 0.089 & 0.922 / 0.695 / 0.459 & 0.922 / 0.939 / 0.922 \\ 
$\beta_1=0.8$ & -0.011 / -0.001 / -0.002 & 0.041 / 0.033 / 0.021 & 0.945 / 0.947 / 0.936 \\ 
$\omega_{x,0}=1.75$ & -0.055 / -0.048 / -0.016 & 0.438 / 0.264 / 0.252 & 0.958 / 0.931 / 0.935 \\ 
$\omega_{x,1}=0.08$ & -0.014 / -0.002 / -0.003 & 0.023 / 0.011 / 0.011 & 0.901 / 0.933 / 0.922 \\ 
$\omega_{y,0}=0$ & 0.097 / -0.015 / 0.012 & 0.693 / 0.639 / 0.330 & 0.923 / 0.945 / 0.944 \\ 
$\omega_{y,1}=0.2$ & -0.025 / -0.023 / -0.006 & 0.041 / 0.038 / 0.016 & 0.824 / 0.868 / 0.917 \\ 
$\text{PoA}$ & 0.030 / 0.023 / 0.007 & 0.034 / 0.027 / 0.011 & 0.703 / 0.807 / 0.962 \\ \hline
\end{tabular}
\caption{Simulation results for Parameter Scenario 4, displayed for Replication Scenarios 1/2/3.}
\label{tab:sim_scen4}
\end{table}

\begin{figure}[H]
    \centering
    \includegraphics[width=1\textwidth]{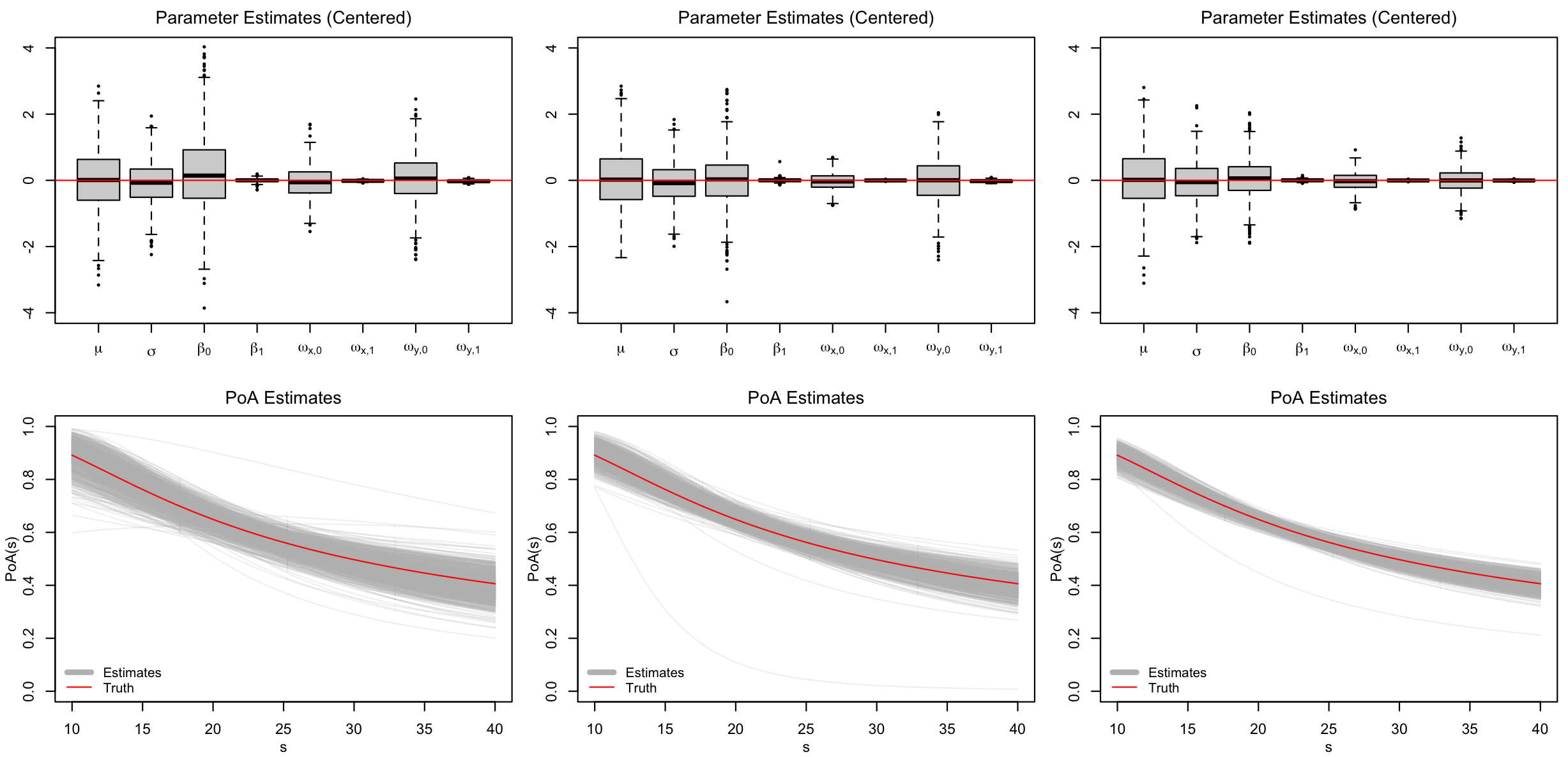}
    \caption{Simulation results for Parameter Scenario 5, displayed for Replication Scenarios 1 (left column), 2 (middle column), and 3 (right columns).}
    \label{fig:sim_scen5}
\end{figure}

\begin{table}[H]
\centering
\begin{tabular}{|l|c|c|c|}
\hline
Parameter & Bias & RMSE & Coverage \\ \hline
$\mu=25$ & 0.021 / -0.004 / 0.013 & 0.912 / 0.877 / 0.893 & 0.947 / 0.942 / 0.940 \\ 
$\sigma=8.66$ & -0.008 / -0.008 / -0.012 & 0.463 / 0.423 / 0.410 & 0.944 / 0.934 / 0.962 \\ 
$\beta_0=4$ & 0.639 / 0.124 / 0.140 & 1.280 / 0.736 / 0.563 & 0.894 / 0.939 / 0.927 \\ 
$\beta_1=1.2$ & -0.021 / -0.003 / -0.004 & 0.055 / 0.035 / 0.025 & 0.929 / 0.944 / 0.947 \\ 
$\omega_{x,0}=1.75$ & -0.046 / -0.041 / -0.041 & 0.478 / 0.262 / 0.251 & 0.947 / 0.933 / 0.941 \\ 
$\omega_{x,1}=0.08$ & -0.015 / -0.002 / -0.002 & 0.025 / 0.011 / 0.011 & 0.869 / 0.927 / 0.940 \\ 
$\omega_{y,0}=0$ & 0.079 / -0.020 / -0.001 & 0.651 / 0.627 / 0.325 & 0.952 / 0.948 / 0.945 \\ 
$\omega_{y,1}=0.2$ & -0.025 / -0.020 / -0.005 & 0.039 / 0.035 / 0.016 & 0.844 / 0.887 / 0.933 \\ 
$\text{PoA}$ & -0.001 / 0.002 / -0.002 & 0.037 / 0.026 / 0.018 & 0.996 / 0.996 / 0.997 \\ \hline
\end{tabular}
\caption{Simulation results for Parameter Scenario 5, displayed for Replication Scenarios 1/2/3.}
\label{tab:sim_scen5}
\end{table}

\begin{figure}[H]
    \centering
    \includegraphics[width=1\textwidth]{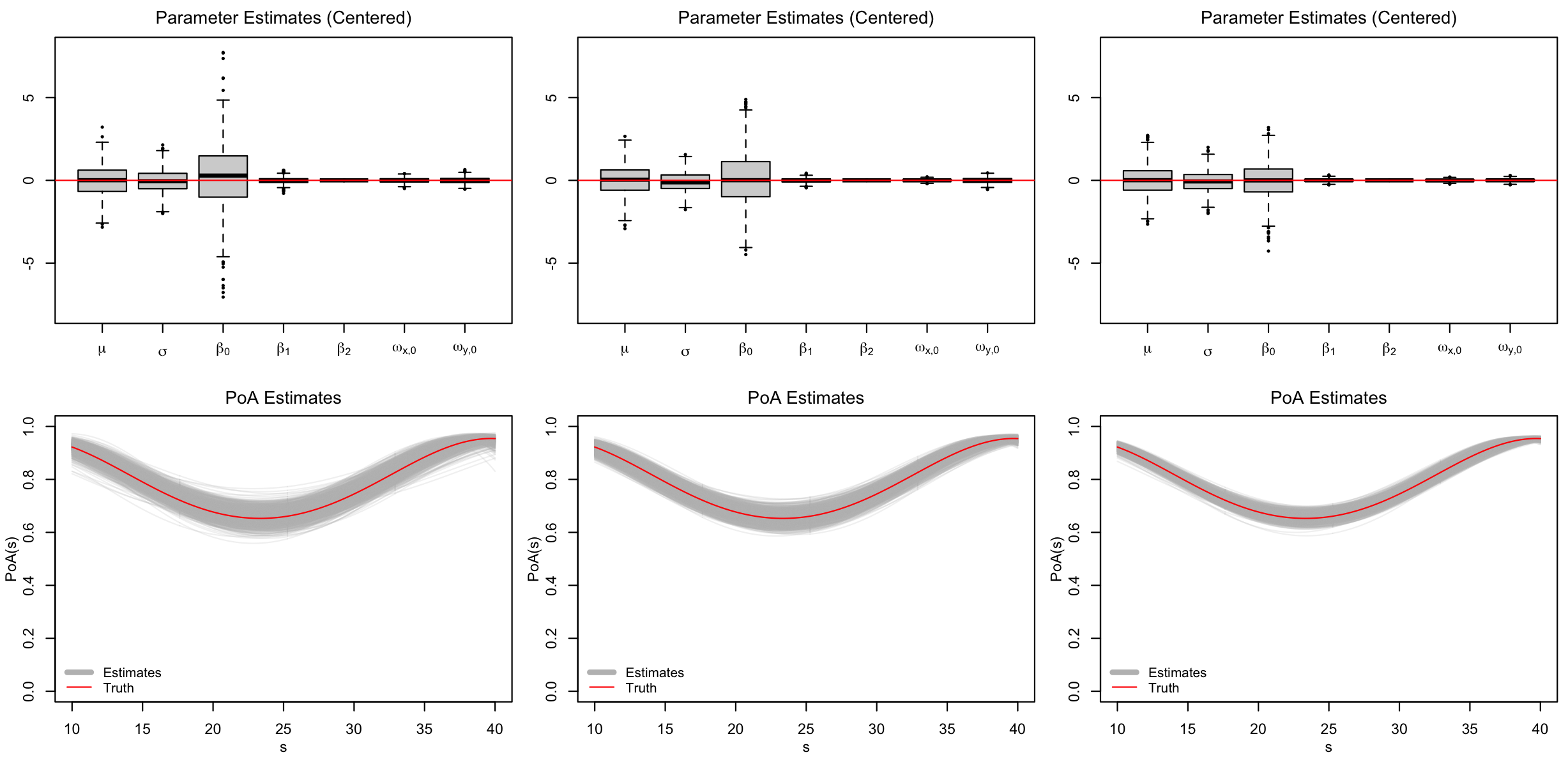}
    \caption{Simulation results for Parameter Scenario 6, displayed for Replication Scenarios 1 (left column), 2 (middle column), and 3 (right columns).}
    \label{fig:sim_scen6}
\end{figure}

\begin{table}[H]
\centering
\begin{tabular}{|l|c|c|c|}
\hline
Parameter & Bias & RMSE & Coverage \\ \hline
$\mu=25$ & -0.029 / -0.043 / -0.012 & 0.917 / 0.883 / 0.870 & 0.943 / 0.937 / 0.945 \\ 
$\sigma=8.66$ & -0.031 / 0.001 / -0.013 & 0.436 / 0.418 / 0.406 & 0.950 / 0.935 / 0.948 \\ 
$\beta_0=8.3$ & -1.468 / -0.492 / -0.591 & 3.009 / 2.204 / 1.557 & 0.914 / 0.930 / 0.927 \\ 
$\beta_1=-0.4$ & 0.137 / 0.047 / 0.056 & 0.275 / 0.195 / 0.143 & 0.903 / 0.927 / 0.930 \\ 
$\beta_2=0.03$ & -0.003 / -0.001 / -0.001 & 0.006 / 0.004 / 0.003 & 0.898 / 0.924 / 0.923 \\ 
$\omega_{x,0}=3$ & -0.003 / -0.001 / -0.004  & 0.140 / 0.071 / 0.068 & 0.934 / 0.948 / 0.950 \\ 
$\omega_{y,0}=4$ & -0.015 / -0.007 / 0.002 & 0.182 / 0.172 / 0.097 & 0.939 / 0.936 / 0.933 \\ 
$\text{PoA}$ & 0.006 / 0.002 / 0.002 & 0.029 / 0.022 / 0.015 & 0.974 / 0.977 / 0.983 \\ \hline
\end{tabular}
\caption{Simulation results for Parameter Scenario 6, displayed for Replication Scenarios 1/2/3.}
\label{tab:sim_scen6}
\end{table}

\begin{figure}[H]
    \centering
    \includegraphics[width=1\textwidth]{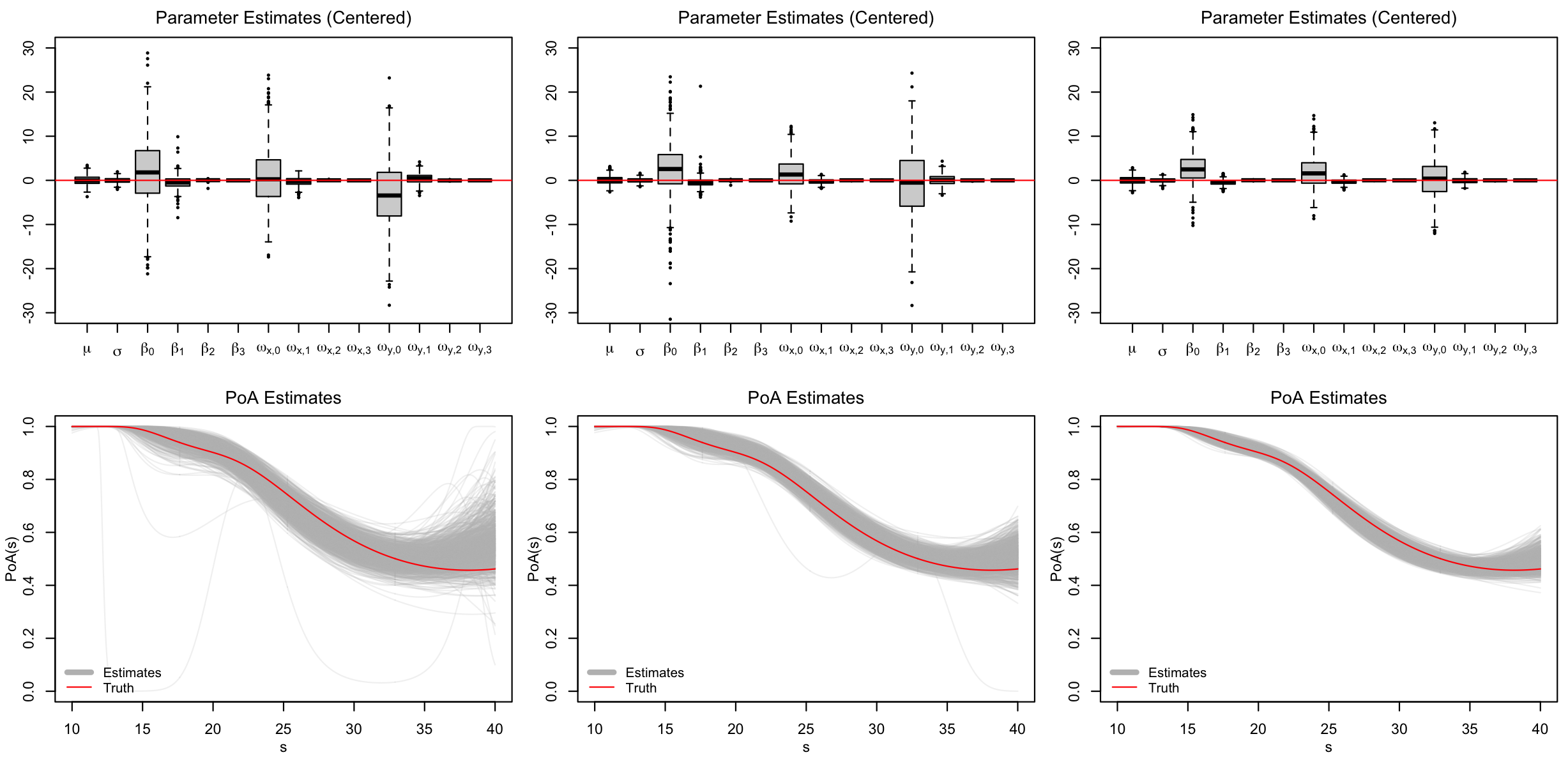}
    \caption{Simulation results for Parameter Scenario 8, displayed for Replication Scenarios 1 (left column), 2 (middle column), and 3 (right columns).}
    \label{fig:sim_scen8}
\end{figure}

\begin{table}[H]
\centering
\begin{tabular}{|l|c|c|c|}
\hline
Parameter & Bias & RMSE & Coverage \\ \hline
$\mu=25$ & 0.065 / -0.017 / 0.002 & 0.989 / 0.895 / 0.912 & 0.953 / 0.940 / 0.937 \\ 
$\sigma=8.66$ & 0.016 / 0.000 / -0.005 & 0.572 / 0.463 / 0.432 & 0.942 / 0.945 / 0.957 \\ 
$\beta_0=20.3125$ & 1.055 / 2.260 / 2.420 & 9.984 / 7.828 / 3.987 & 0.959 / 0.960 / 0.903 \\ 
$\beta_1=-2.9375$ & -0.366 / -0.437 / -0.465 & 1.639 / 1.312 / 0.703 & 0.954 / 0.950 / 0.887 \\ 
$\beta_2=0.1875$ & 0.031 / 0.026 / 0.027 & 0.083 / 0.066 / 0.038 & 0.948 / 0.944 / 0.861 \\ 
$\beta_3=-0.0025$ & -0.001 / -0.000 / -0.000  & 0.001 / 0.001 / 0.001 & 0.933 / 0.939 / 0.849 \\ 
$\omega_{x,0}=5.4$ & -0.840 / 1.730 / 1.510  & 6.884 / 3.704 / 3.641 & 0.937 / 0.920 / 0.924 \\ 
$\omega_{x,1}=-1$ & -0.054 / -0.302 / -0.272 & 0.997 / 0.577 / 0.563 & 0.934 / 0.909 / 0.914 \\ 
$\omega_{x,2}=0.06$ & 0.012 / 0.016 / 0.015 & 0.046 / 0.028 / 0.027 & 0.927 / 0.891 / 0.897 \\ 
$\omega_{x,3}=-0.0008$ & -0.000 / -0.000 / -0.000 & 0.001 / 0.000 / 0.000 & 0.899 / 0.871 / 0.880 \\ 
$\omega_{y,0}=5.44$ & -4.020 / -0.631 / -0.175 & 8.396 / 7.981 / 3.998 & 0.926 / 0.949 / 0.957 \\ 
$\omega_{y,1}=-0.98$ & 0.519 / 0.082 / -0.010 & 1.193 / 1.191 / 0.595 & 0.924 / 0.942 / 0.951 \\ 
$\omega_{y,2}=0.062$ & -0.019 / -0.004 / 0.002 & 0.051 / 0.054 / 0.027 & 0.932 / 0.941 / 0.950 \\ 
$\omega_{y,3}=-0.00083$ & 0.000 / 0.000 / -0.000 & 0.001 / 0.001 / 0.000 & 0.937 / 0.942 / 0.950 \\ 
$\text{PoA}$ & 0.008 / 0.013 / -0.002 & 0.042 / 0.030 / 0.017 & 0.454 / 0.711 / 0.897 \\ \hline
\end{tabular}
\caption{Simulation results for Parameter Scenario 8, displayed for Replication Scenarios 1/2/3.}
\label{tab:sim_scen8}
\end{table}

\newpage

\bibliography{references}

\end{document}